\documentclass{aa}
\usepackage{etoolbox}

\usepackage{graphicx}
\usepackage{txfonts}
\usepackage{float}
\usepackage{hyperref}
\hypersetup{
     colorlinks = true,
     linkcolor = blue,
     anchorcolor = blue,
     citecolor = blue,
     filecolor = blue,
     urlcolor = blue
     }
\usepackage{pdflscape} 
\usepackage{afterpage} 

\usepackage{tabularx}
\usepackage{breqn}
\usepackage{geometry}
\usepackage{lipsum}

\makeatletter
\renewcommand*\aa@pageof{, page \thepage{} of \pageref*{LastPage}}
\usepackage{caption}
\usepackage{multirow}
\usepackage{lscape}
\usepackage{orcidlink}
\usepackage{float}

\begin{document} 

   \title{
   Kinematic scaling relations of disc galaxies from ionised gas at $z\sim~1$ and their connection with dark matter haloes} 
   \titlerunning{Ionised gas kinematics of $z\sim1$ discs and their connection with dark matter haloes}
\authorrunning{Mancera Pi\~na et al.}

   \author{Pavel E. Mancera Pi\~na\inst{1}\fnmsep\thanks{\email{pavel@strw.leidenuniv.nl}}\orcidlink{0000-0001-5175-939X}, 
   Enrico M. Di Teodoro\inst{2}\orcidlink{0000-0003-4019-0673},
 S. Michael Fall\inst{3}\orcidlink{0000-0003-3323-9061},
Antonino Marasco\inst{4}\orcidlink{0000-0002-5655-6054},\\
   Mariska Kriek\inst{1}\orcidlink{0000-0002-7613-9872},
   and Marco Martorano\inst{5}\orcidlink{0000-0003-2373-0404}. 
          }
   \institute{Leiden Observatory, Leiden University, P.O. Box 9513, 2300 RA, Leiden, The Netherlands 
   \and Dipartimento di Fisica e Astronomia, Università degli Studi di Firenze, via G. Sansone 1, 50019 Sesto Fiorentino, Firenze, Italy
   \and Department of Physics and Astronomy, Johns Hopkins University, 3400 N. Charles Street, Baltimore, MD 21218, USA
   \and INAF - Padova Astronomical Observatory, Vicolo dell'Osservatorio 5, I-35122 Padova, Italy
   \and Sterrenkundig Observatorium, Universiteit Gent, Krijgslaan 281 S9, 9000 Gent, Belgium
          }

   \date{}

  \abstract{
We derive the Tully-Fisher (TFR; $M_\ast-V_{\rm circ,f}$) and Fall (FR; $j_\ast-M_\ast$) relations at redshift $z = 0.9$ using a sample of 43 main-sequence disc galaxies with H$\alpha$ IFU data and JWST/HST imaging. The strength of our analysis lies in the use of state-of-the-art 3D kinematic models to infer galaxy rotation curves, the inclusion of near-IR bands and their morphological modelling, and the application of homogeneous spectral energy distribution modelling to our photometry measurements to estimate stellar masses. After correcting the inferred H$\alpha$ velocities for asymmetric drift, we find a TFR of the form $\log(M_\ast / M_\odot) = a \log(V_{\rm circ,f} / 150~\mathrm{km\,s^{-1}}) + b$, with $a=3.82^{+0.55}_{-0.40}$ and $b=10.27^{+0.06}_{-0.07}$, as well as a FR of the form $\log(j_\ast / \mathrm{kpc\,km\,s^{-1}}) = a \log(M_\ast / 10^{10.5} M_\odot) + b$, with $a=0.44^{+0.06}_{-0.06}$ and $b=2.86^{+0.02}_{-0.02}$.

\looseness=-1
Compared with their $z=0$ counterparts, we found moderate evolution in the TFR and strong evolution in the FR over the past $8$ Gyr. We interpreted our findings in the context of the galaxy-to-halo scaling parameters $f_{\rm M}=M_\ast/M_{\rm vir}$ and $f_{\rm j}=j_\ast/j_{\rm vir}$. 
We inferred that $f_{\rm j}$ shows little redshift evolution and depends very weakly on $M_\ast$, with typical values around $f_{\rm j}\sim0.8$. 
As for $f_{\rm M}$, we find it to be higher and less dependent on $M_\ast$ at $z=0.9$ than at $z=0$. We discuss how interpreting our observed $f_{\rm M}-M_\ast$ relations within the cold dark matter framework implies necessarily that the galaxy populations at $z=0.9$ and $z=0$ are not the progenitor nor descendant of one another. The alternative scenario is that the $z=0.9$ scaling relations are incorrect due to strong selection effects, unidentified systematics, or the possibility that H$\alpha$ kinematics may not be a reliable dynamical tracer. Such problems would affect not only our work but also previous studies on the same subject.}


   \maketitle

\section{Introduction}
\label{sec:intro}

Two of the most fundamental scaling relations for disc galaxies are those connecting their stellar mass ($M_\ast$) with their rotational velocities ($V$) and their stellar specific angular momentum ($j_\ast$), i.e. the stellar Tully-Fisher (TFR; \citealt{tullyfisher,bell2001_M2L}) and Fall (FR; \citealt{fall83}) relations.
These scaling relations are closely intertwined with the process of galaxy evolution. According to our current galaxy formation theories, dark matter and baryons acquire their angular momentum through tidal torques from neighbouring systems before virialisation. Upon the dissipation and gravitational collapse of the baryons, the global angular momentum is approximately conserved, leading the baryons in disc galaxies to settle into relatively thin disc-like structures supported by rotation and embedded in more extended dark matter haloes without significant rotational support (e.g. \citealt{peebles,binney1977,white1978,fall1980,blumenthal1984,white1984,dalcanton_discs,mo98}). In this way, the distribution of angular momentum is thought to set the rotational velocity and mass distribution of galaxies, regulating also their morphology and gas content (e.g. \citealt{romanowsky,OG14,cortese2016,lagos_eagle,swinbank2017,sweet2020,paperIIBFR,nikki}). Observational constraints on the TFR and FR provide crucial benchmarks for galaxy formation models.

In the nearby Universe, at $z\sim0$, the TFR and FR have been studied in detail and found to be well described by unbroken power laws of the form $M_\ast \propto V^a$ and $j_\ast\propto M_\ast^a$, respectively. These slopes are found to be $a\approx3.5-5$ for the TFR (e.g. \citealt{bell2001_M2L,reyes2011,anastasiaphotometry,catinella2023,marasco_mstar}) and $a\approx0.5-0.6$ for the FR (e.g. \citealt{ postijstar,paperIBFR,hardwick2022,superspirals_enrico,enrico_massmodels_ss,marasco_mstar}). At higher redshifts, the situation remains significantly less certain. Although different works have studied the TFR and FR up to $z\sim2.5$ (e.g. \citealt{conselice2005,cresci2009,burkert_j,contini2016,enrico_z1,price2016,kross_harrison,swinbank2017,gillman2019,marascojstar,sweet2019,pelliccia2019,tiley2019,mercier2023}), there is no consensus on whether or not the relations (slopes, intercept, intrinsic scatter) evolve with $z$ (see \citealt{ubler2017,pelliccia2019,bouche2021,sharma2024,espejo2025}). Some of the difficulty in tracing the evolution of the TFR and FR arises from observational uncertainties and methodological limitations when studying high-$z$ disc galaxies (e.g. \citealt{rizzo2022,araujocarvalho2025,sharma2025}). Two major concerns in most previous works are the derivation of galaxy kinematics and of stellar masses. Obtaining reliable kinematics at high $z$ is challenging, due to both the limited quality of the data (low spatial resolution, low signal-to-noise ratio, short radial extent) and the intrinsically complex structure of young galaxies.
In addition, $M_\ast$ is typically estimated through spectral energy distribution (SED) fitting (or by adopting a fixed mass-to-light ratio), often relying on photometry of variable quality and lacking rest-frame near-infrared (NIR) coverage, which is crucial for constraining the underlying stellar mass distribution.

In this work, we revisit the evolution of the TFR and FR at $z\sim1$ while addressing these limitations. In particular, our study improves upon previous efforts in several key ways: \emph{i)} We use well-tested software that takes into account observational effects for data of limited quality in a self-consistent way. \emph{ii)} We exploit new JWST observations providing exquisite NIR data, improving the stellar mass density profiles. \emph{iii)} We apply asymmetric drift corrections to take into account the pressure-supported motions of ionised gas and stars compared to cold gas (which is typically used at $z=0$); this is crucial as it is a systematic effect that depends on $M_\ast$. \emph{iv)} Rather than assuming stellar-to-halo mass and specific angular momentum ratios that depend on abundance-matching calibrations, we show that these can be directly derived from the observed TFR and FR. \emph{v)} Finally, we assess the evolution of the TFR and FR at $z\sim1$ by contrasting against new and improved $z=0$ determinations \citep{marasco_mstar}. Throughout this work, we adopt a $\Lambda$ cold dark matter ($\Lambda$CDM) cosmology with $\Omega_{\mathrm{m},0} = 0.3$, $\Omega_{\Lambda,0} = 0.7$, and $H_0 = 70~\rm{km\,s^{-1}\,Mpc^{-1}}$.

\section{Data and sample selection}
\label{sec:data}
For this study, we made use of data tracing the resolved stellar mass distribution and kinematics of galaxies at $z\approx1$. For the kinematics, we relied on the integral field unit (IFU) spectroscopy from the KROSS \citep{kross} and KMOS$^{\rm 3D}$ \citep{kmos3d} surveys. These surveys mapped the H$\alpha$ distribution and kinematics of galaxies at $z\sim0.6-2.5$ using the KMOS instrument at the VLT. The spectral resolution of the data is $R=\lambda/\Delta\lambda \sim 3500-4000$, which for H$\alpha$ at our $z$ corresponds to $\sigma\sim40$ km/s. The typical full width at half maximum (FWHM) of the point-spread function (PSF) is $\sim0.5-0.8$ arcsec \citep{kross,kmos3d_final}. We note that the kinematics of these samples have been studied before (e.g., \citealt{kross,burkert_j,kross_harrison,ubler2017,tiley2019,sharma2021}), although with approaches different from ours, as discussed below.

For this work, we required a galaxy sample that met several selection criteria on the kinematic data. We started by visually inspecting all the position-velocity (PV) slices along the major and minor axes of the galaxies in the parent sample, keeping only those systems with a high signal-to-noise (S/N) and compelling regular kinematics (i.e. clear velocity gradients in the major-axis PV, no gradients in the minor-axis PV, no signs of mergers) to ensure a clean sample of rotating discs. Next, we selected against poorly resolved galaxies with H$\alpha$ emission less extended than $\sim2$ times the PSF. As shown below, we verified that this criterion does not appear to significantly bias the optical mass-size relation of our galaxies, which is consistent with literature relations for the star-forming disc population.

In addition, we require that the galaxies have available HST- or JWST-based morphological Sérsic models. Specifically, we rely on the results from \cite{vanderwel2012,martorano2024}, and \cite{martorano_2025}, who carefully built PSFs and used \textsc{galfit} (\citealt{galfit}, see \citealt{martorano2023} for the treatment of the uncertainties) to obtain accurate effective radii ($R_{\rm eff,\ast}$), Sérsic indices ($n$), position angles, and axial ratios ($b/a$).  
We prioritised the parameters obtained from the reddest available band, which is f444w from NIRCam on JWST for $60\%$ of our final sample (see below) and f160w from WFC3 on HST for the remaining $40\%$. These parameters and our $M_\ast$ estimates (described next) define $\Sigma_\ast(R)$, our stellar-mass surface-density profiles. We note that at our redshifts, f160w and f444w trace rest-frame emission at $0.81\,\mu$m and $2.33\,\mu$m, respectively, roughly corresponding to the rest-frame $I-$ and $K-$bands. At our redshift and $M_\ast$ range, galaxy sizes in these bands agree within $10-20\%$ \citep{vanderwel2014,suess2022}.

\looseness=-1
A key aspect of this study is the use of robust $M_\ast$ estimates, which we obtained using the SED fitting approach by \cite{marasco_mstar} and exploiting our HST (ACS or WFC3) and/or JWST (NIRCam) photometry. In particular, we used the publicly available images from the following surveys: CANDELS \citep{candels1,candels2}, COSMOS \citep{cosmos1,cosmos2}, PRIMER \citep{primer2021}, JADES \citep{jades}, and COSMOS-Web \citep{cosmosweb}. These surveys provide imaging in (some of) the following filters: f444w, f356w, f277w, f150w, f115w, and f090w from JWST, and f160w, f125w, f105w, f814w, f850lp, and f606w from HST. Using only space-based imaging (a condition not always imposed in previous studies) is crucial, as it has significantly better spatial resolution, photometric calibration, and overall image quality than ground-based data. Moreover, unlike previous studies, we used JWST rest-frame bands, enabling us to better characterise the stellar masses of our sample. We carefully extracted the photometry (see \citealt{marasco_mstar} for full details) in the available bands and modelled the SED with the software \textsc{Bagpipes} \citep{bagpipes}. We adopted `non-parametric' star-formation histories, considering the stellar population models from \citet[][2016 version]{bc03}\footnote{We also explored the BPASS stellar population models \citep{eldridge2009,stanway2018}, finding that they produce stellar masses lower 0.25 dex on average but without significant systematics, and therefore do not affect the slopes of the scaling relations presented in Sect.~\ref{sec:scaling}.}, the dust model from \cite{charlot2000}, and a \cite{kroupa2002} initial mass function. The redshift is fixed to the H$\alpha$ value from the KROSS and KMOS$^{\rm 3D}$ surveys. We only considered galaxies with fluxes in at least four bands (we typically have five or six bands but as many as ten) and whose model SEDs provide a high-quality fit to the data ($p(>\chi^2)>0.05$; see \citealt{marasco_mstar}). \cite{marasco_mstar} showed that their SED-based and dynamical $M_\ast$ estimates in local disc galaxies (e.g. \citealt{postinomissing,schombert2022,paper_galaxyhalo})
agree within $\sim0.2$ dex across nearly six orders of magnitude in $M_\ast$, which implies that they can be used interchangeably in the study of scaling relations and their evolution.

\begin{figure*}
    \centering
    \includegraphics[width=1\linewidth]{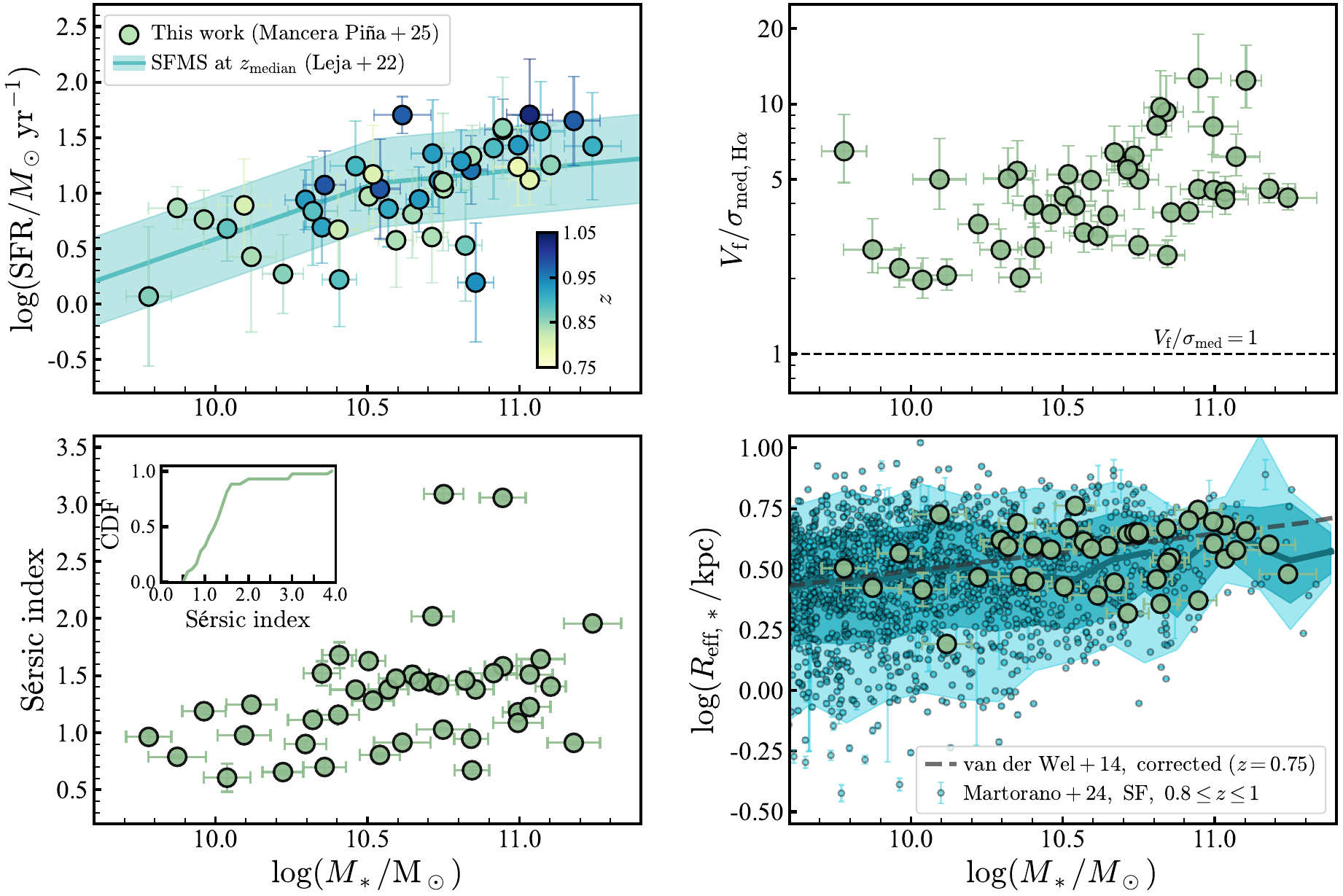}
    \caption{\textit{Top left:} The SFMS defined by our high-$z$ galaxies, colour-coded by their redshift. The reference SFMS from \cite{leja2022} is shown. Star formation rates come from our SED fitting.
    \textit{Top right:} Rotation to dispersion ratio for our galaxy sample. 
    \textit{Bottom left:} Sérsic index as a function of $M_\ast$. The inset shows the cumulative distribution of the Sérsic indices. \textit{Bottom right:} Stellar mass-size relation. Our galaxies (green markers) are contrasted against 1) the individual measurements from \cite{martorano2024} at $0.8 \leq z \leq 1$ (blue markers) and their $1$ and $2\,\sigma$ ranges (blue bands), and 2) the relation from \cite{vanderwel2014} at $z=0.75$ after applying approximate corrections to account for the sizes differences in optical vs. NIR bands.}
    \label{fig:sample}
\end{figure*}

Assuming an intrinsic disc axial ratio $q_0=0.2$ (e.g. \citealt{fouque1990}), we derived galaxy inclinations from the observed $b/a$ using the standard \citep{hubble1926} relation $\cos^2(i) = ((b/a)^2 - q_0^2)/(1 - q_0^2)$. We excluded galaxies with $i < 30^\circ$ (where small errors in $i$ cause significant velocity uncertainties) and $i > 80^\circ$ (for which ring-by-ring kinematic models are not well suited due to the overlap of different lines of sight). From the remaining sample, we retained only systems for which 3D kinematics can be reliably modelled. For this, we used the forward modelling software $^{\rm 3D}$Barolo \citep{barolo}, which accounts for beam smearing and enables robust recovery of intrinsic rotation curves and velocity dispersions. Full details of the kinematic modelling are provided in Appendix~\ref{app:kinmodels}.

After applying our selection criteria, we selected 43 disc galaxies. Our sample spans the range $0.78 < z < 1.03$ with a median $z$ of $0.89$. This redshift range corresponds to cosmic ages between $6.7\,\rm{Gyr}$ and $5.6\,\rm{Gyr}$. The galaxies have stellar masses in the range $5\times10^9 < M_\ast/M_\odot < 2\times10^{11}$.
Given all our selection cuts based on data quality, establishing a selection function is far from straightforward. However, we found that our final sample is representative of the typical star-forming galaxy population, characterised by regular kinematics. This is illustrated in Fig.~\ref{fig:sample}, showing that the sample follows the star formation main sequence (SFMS, top left panel) at $z\approx0.9$, is dominated by rotation\footnote{Our median rotation-to-dispersion ratio $V_{\rm f}/\sigma_{\mathrm{H}\alpha,\mathrm{med}}$ is close to 5 (see Appendix~\ref{app:kinmodels}), where $\sigma_{\mathrm{H}\alpha,\mathrm{med}}$ is the median H$\alpha$ velocity dispersion radial profile. Since $ \sigma_{\mathrm{H}\alpha}$ is typically twice that of the cold gas \citep{levy2018,girard2021,ejdetjarn2022}, this correction would align \( V_{\rm f}/\sigma_{\mathrm{H}\alpha,\mathrm{med}} \) with typical values observed at \( z = 0 \) in disc galaxies with cold gas kinematics (e.g. \citealt{paper_galaxyhalo}).
} (top right panel), has a Sérsic index distribution typical of discs (median $n\approx1.3$, bottom left panel), and follows the $M_\ast-R_{\rm eff,\ast}$ relation from \cite{martorano2024} at $z=0.9$ and \cite{vanderwel2014} at $z=0.75$ (bottom right panel\footnote{We have modified the relation from \cite{vanderwel2014} to account for typical differences between their optical sizes and NIR sizes (used in \citealt{martorano2024} and this work). Specifically, we rescaled the optical sizes by linear interpolation so that they decrease by 0.1 dex at $\log(M_\ast/M_\odot)=9.5$ and by 0.22 dex at $\log(M_\ast/M_\odot)=11.3$, consistent with \citet[][see their Fig.~7]{martorano2024}.}). Table~\ref{tab:cat} lists the main properties of our sample.

\section{Estimating $V_{\rm circ,f}$ and $j_\ast$}
\label{sec:method}
Our kinematic modelling described in Appendix~\ref{app:kinmodels} allows us to obtain the intrinsic rotational velocity of the ionised gas ($V_{{\rm H}\alpha}(R)$) and its velocity dispersion ($\sigma_{{\mathrm H}\alpha}(R)$). Instead, the TFR requires the circular speed ($V_{\rm circ}$\footnote{$V_{\rm circ}(R)$ is the speed of a test particle in pure circular orbit at a radius $R$, e.g. \cite{binney, bookFilippo}. $V_{\rm circ}$ is close to the rotational speed of H\,{\sc{i}}, typically used at $z=0$ to derive the TFR.}), and the FR requires the stellar rotational velocities ($V_\ast$). If derived from cold gas (e.g. H\,{\sc{i}} or CO), the rotational velocities could be used directly to build the TFR and FR, since the cold gas has a high rotational support and low pressure-supported motions, and the rotational speeds are close to both $V_{\rm circ}$ and $V_\ast$ (at least for $M_\ast\gtrsim5\times10^8\,M_\odot$, see \citealt{paperIBFR}). However, for ionised gas at $z\sim1$ (see also \citealt{catinella2023} at $z=0$), the velocity dispersions tend to be higher and the rotation-to-dispersion ratios lower, making it imperative to correct for the different asymmetric drift of the stars and ionised gas. In particular, $V_{\rm circ}$, $V_\ast$, and our measured $V_{{\rm H}\alpha}$ are related through the expressions
\begin{equation}
    V_{\rm circ}^2 = V_{\mathrm{H} \alpha}^2 + V_{\mathrm{AD,H}\alpha}^2~,
\end{equation} 
\begin{equation}
    V_\ast^2 = V_{\rm circ}^2 - V_{\rm AD,\ast}^2~,
\end{equation}
where $V_{\rm AD,\ast}$ and $V_{\mathrm{AD,H} \alpha}$ are the asymmetric drift (AD) corrections to account for the pressure support provided by the stars and the ionised gas, respectively. Therefore, the steps to follow are first to convert our $V_{{\rm H}\alpha}$ into $V_{\rm circ}$, and then $V_{\rm circ}$ into $V_\ast$. We note that in this study, all uncertainties are propagated throughout our calculations using Monte Carlo sampling (with measurement errors assumed to be Gaussian). The quoted errors correspond to the difference between the 84th and 16th percentiles and are propagated using the full sampling of the posterior distributions.

Under the assumptions of constant scale heights and relatively thin discs, the AD corrections \citep{binney} take the approximate form 
\begin{equation}
    V_{\mathrm{AD}}^2(R) = -R \left( \frac{\sigma_z(R)}{\beta} \right)^2\ \frac{\partial \ln \left( \Sigma(R)\, \sigma_z(R)^2 \right)}{\partial R}\ ~,
\end{equation}
with $\Sigma$ the ionised gas or stellar surface density, and $\beta\equiv\sigma_z/\sigma_R$, with $\sigma_z$ and $\sigma_R$ the vertical and radial components of the stellar or ionised gas velocity dispersion profiles.

\looseness=-3
We start by computing $V_{\mathrm{AD,H}\alpha}$. For this, we make the common assumption that the gas velocity dispersion is isotropic, which sets $\beta=1$. Next we assume a constant $\sigma_{z,{\rm H}\alpha}(R)$, given by the median value of our observed $\sigma_{\mathrm{H}\alpha}$\footnote{This was done for simplicity and to avoid over-interpreting the data. For some galaxies, $\sigma_{\mathrm{H}\alpha}(R)$ may not be smooth and shows discontinuities driven by the limited S/N and resolution of the observations. In any case, we corroborated that using the actual $\sigma_{\mathrm{H}\alpha}(R)$ profile does not significantly affect the results presented below.}. Regarding the surface density, although in principle we have access to $\Sigma_{\mathrm{H}\alpha}$, the H$\alpha$ data's spatial resolution is too low to infer detailed radial profiles. Instead, we assume that H$\alpha$ shares the same Sérsic index as the stellar disc, but with $R_{\mathrm{eff,H}\alpha} = 1.13(\pm0.05)\,R_{\rm eff,\ast}$, as found for disc galaxies within our $M_\ast$ and $z$ range \citep{nelson2016,wilman2020}. With this $V_{\mathrm{AD,H}\alpha}$, we can convert $V_{{\rm H}\alpha}$ into $V_{\rm circ}$.

Kinematic data at $z\sim0$ have extended H\,{\sc{i}} data that can be traced far out the disc with many resolution elements to measure the characteristic flat velocity of $V_{\rm circ}(R)$ ($V_{\rm circ,f}$), which we need for the TFR. In contrast, high-$z$ rotation curves have more limited resolution (typically three to five (nearly) independent resolution elements, see Appendix~\ref{app:kinmodels}) and do not extend as much as H\,{\sc{i}}, which prevents us from performing the same measurement. Instead, we implement the following approach. We first fit $V_{\rm circ}(R)$ with the arctan rotation curve model from \cite{courteau2007}, i.e. $V_{\rm circ}(R) = (2/\pi)\,V_{\rm a}\,\arctan(R/r_{\rm t})$, with $V_{\rm a}$ the asymptotic velocity of the rotation curve and $r_{\rm t}$ a turnover radius between the rising and outer part of the rotation curve. In practice, the fit is done using the Bayesian Nested Sampling software \texttt{dynesty} \citep{dynesty}, minimising a $\chi^2$ likelihood. From empirical tests using a sample of $z=0$ disc galaxies with high-resolution kinematics (described in Appendix~\ref{app:sparc}), we find that we can estimate $V_{\rm circ,f}$ with a $\sim1\%$ accuracy, by evaluating the best-fit arctan model at $R=2\,R_{\rm eff,\ast}$ for galaxies with $V_{\rm a} \geq 100\,\rm{km/s}$, (as those in our high-$z$ sample), and at $R=3\,R_{\rm eff,\ast}$ for lower $V_{\rm a}$. From this, we obtain the $V_{\rm circ,f}$ needed for the TFR.

Next, we focus on $V_{\rm AD,\ast}$, which allows us to estimate $V_\ast$ for the FR. We follow the procedure detailed in \cite{paperIBFR}. Specifically, as supported by theoretical and observational work at $z=0$ \citep{vanderkruit,martinssonVI}, we consider that the stellar velocity dispersion profile has a radial profile of the form
\begin{equation}
    \sigma_{z,\ast}(R)/\mathrm{km\,s^{-1}} = \mathrm{max}[\sigma_{z_{0,\ast}}\exp(-R/1.19R_{\rm eff,\ast}),\ 10]~,
\end{equation}
with an empirical $\sigma_{z_{0,\ast}}$ derived by \cite{paperIBFR} based on measurements of disc galaxies at $z=0$ (see also e.g. \citealt{martinssonVI}) given by
\begin{equation} 
    \dfrac{\sigma_{z_0,\ast}}{\rm{km\,s}^{-1}}  = 9.7 \left( \dfrac{V_{\rm circ,f}}{100\,\rm{km\,s}^{-1}}\right) ^2 + 2.6 \left(\dfrac{V_{\rm circ,f}}{100\,\rm{km\,s^{-1}}}\right) + 10.61(\pm5)~~,
\end{equation}
which we assume not to evolve with $z$. The above calibrations are semi-empirical, but it is worth noting that detailed measurements show some degree of diversity among galaxies \citep{mogotsi2019}. Finally, we consider $\beta=0.8\pm0.2$, consistent with the values found for nearby disc galaxies by \cite{mogotsi2019}. With all of the above, we compute $V_{\rm AD,\ast}$ and convert $V_{\rm circ}$ into $V_\ast$. \vskip 4pt

With $V_\ast$, we can determine $j_\ast$, which is defined as
\begin{equation}
j_\ast = \dfrac{2\pi \int {R'}^{2} ~\Sigma_\ast(R')~V_\ast(R')~dR'}{2\pi \int  {R'}~\Sigma_\ast(R')~dR'}~.
    \label{eq:j}
\end{equation}
For $z=0$ data, Eq.~\ref{eq:j} is computed by summing the high-resolution rotation curves and mass profiles (e.g. \citealt{postijstar,paperIBFR,enrico_massmodels_ss}). For the $z=0.9$ data, we rely on our Sérsic models for $\Sigma_\ast$ and on new arctan fits for $V_\ast(R)$. We integrate Eq.~\ref{eq:j} (using deprojected parameters, considering the inclination angle) up to $R_{\rm out}=10\, R_{\rm{eff}}$, ensuring the convergence of $j_\ast$. As detailed in Appendix~\ref{app:sparc}, we have tested our procedure on a comparison sample at $z=0$, finding that $j_\ast$ is recovered with a $\sim5\%$ accuracy.

Before delving into the best-fitting TFR and FR implied by our measurements, we emphasise the importance of applying AD corrections when analysing ionised gas kinematics, which is often overlooked in studies of the TFR and FR at high-$z$. As mentioned before, using such corrections is key to account for the varying rotation-to-dispersion ratios arising from various kinematic tracers at different redshifts, which are also a function of $M_\ast$ since typically low-mass galaxies have less rotational support (Fig.~\ref{fig:sample}). To illustrate this, we quote the average differences in $\log(V_{\rm circ,f})$ and $\log(j_\ast)$ for our galaxy sample implementing or not the AD corrections. At $\log(M_\ast/M_\odot)=10, 10.5$, and $11$, the AD-corrected values for $\log(V_{\rm circ,f})$ are typically higher by about 0.09 dex, 0.04 dex, and 0.02 dex, respectively. Similarly, at the same masses, the AD-corrected values for $\log(j_\ast)$ are higher by about 0.05 dex, 0.02 dex, and 0.0 dex, on average. As we discuss in the next section, these offsets are large enough to impact the best-fitting TFR and FR.

\section{Scaling relations}
\subsection{Scaling laws at $z=0.9$}
\label{sec:scaling}
\begin{figure*}
    \centering
    \includegraphics[width=1\linewidth]{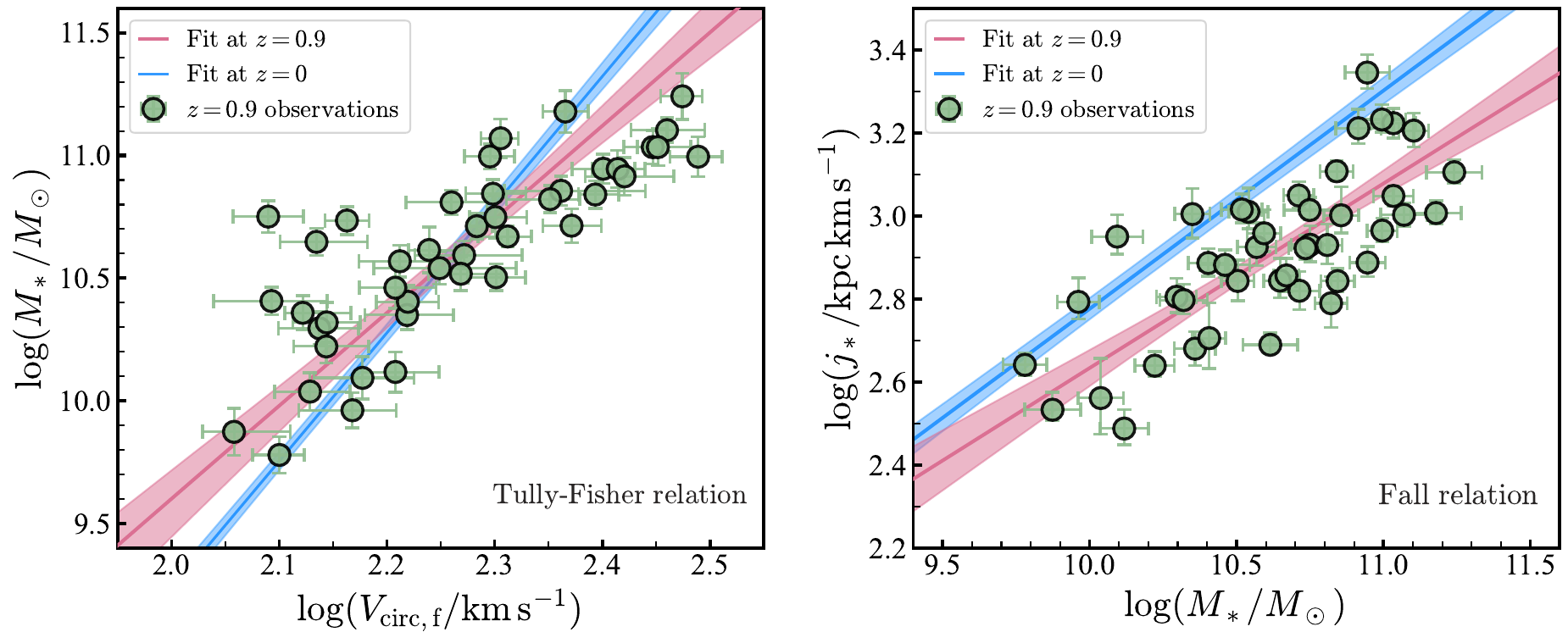}
    \caption{Our $z=0.9$ scaling laws. The TFR is shown on the \emph{left}, and the FR on the \emph{right}. Our measurements are shown with the green markers, while the best-fit relations and their $1\,\sigma$ confidence bands are shown as solid pink curves and bands, respectively. For comparison, we show the $z=0$ TFR and FR from \cite{marasco_mstar}. 
    }
    \label{fig:relationsz1}
\end{figure*}

In Fig.~\ref{fig:relationsz1}, we show our measurements in the $M_\ast-V_{\rm circ,f}$ and $j_\ast-M_\ast$ planes. As found at low-$z$, our high-$z$ galaxies follow well-defined sequences of increasing $V_{\rm circ,f}$ and $j_\ast$ with increasing $M_\ast$. In Fig~\ref{fig:j_vsigma}, we show that at fixed $M_\ast$ galaxies with a high $V_{\rm f}/\sigma_{\rm med}$ also have a high $V_{\rm circ,f}$ and $j_\ast$. We parametrise the observed distributions in the $M_\ast-V_{\rm circ,f}$ and $j_\ast-M_\ast$ planes with power-law models of the form
\begin{equation}
\label{eq:vfit}
    \log \left( \dfrac{M_\ast}{M_\odot} \right) = a \log\left( \dfrac{V_{\rm{circ,f}}}{150~\rm{km\,s^{-1}}} \right) + b~,
\end{equation}
\begin{equation}
\label{eq:jfit}
\log \left( \dfrac{j_\ast}{\mathrm{kpc\,km\,s^{-1}}} \right) = a \log\left( \dfrac{M_\ast}{10^{10.5} M_\odot} \right) + b~,
\end{equation}

\noindent
where $a$ is the slope of the relation and $b$ is the intercept at our pivot mass and velocity. The pivot values of $10^{10.5}\,M_\odot$ and $150\,\rm{km/s}$, chosen to be relatively close to the median values for our high-$z$ sample, help in reducing the covariance between slopes and intercepts.  
In practice, the power-law fits to our TFR and FR are obtained using \texttt{dynesty}, adding an extra term to allow for intrinsic scatter perpendicular to the best-fit relation ($\epsilon_\perp$, the scatter unaccounted for by the observational uncertainties and therefore assumed to be intrinsic to the relations), and including uncertainties in both variables. We use flat priors for $a$, and $b$, and $\epsilon_\perp \geq0$ for $\epsilon_\perp$. The best-fitting relations are shown in Fig.~\ref{fig:relationsz1}, and the best-fitting coefficients are listed in Table~\ref{tab:coeff_z1}, where we also quote the observed vertical RMS scatter of the relations ($\sigma_{M_\ast}$ and $\sigma_{j_\ast}$). We emphasise that $\sigma_{M_\ast}$ and $\sigma_{j_\ast}$ measure the scatter of the data, but do not weight the observational errors, while $\epsilon_\perp$ does, and it is assumed to be inherent to the scaling laws.

The literature on the TFR and FR at $z\sim1$ is extensive, as recently summarised in \cite{sharma2024} and \cite{espejo2025}. Formally, the slope of our TFR is higher than the value of $a=3.03$ by \cite{sharma2024} and similar to those ($a\sim3.6-3.8$) reported by \cite{enrico_z1,kross_harrison,ubler2017,pelliccia2019,tiley2019,melgarejo2021}, although it should be kept in mind the usage of different velocity conventions when defining the TFR, as well as the fact that some of the above studies fixed the slope to the local value from \cite{reyes2011}. Our FR slope is slightly below the $a\sim0.5-0.6$ reported by \cite{kross_harrison,gillman2020} and \cite{bouche2021}, but entirely consistent within the uncertainties.
Comparing the intercepts is more challenging because different studies use different pivot points; however, a visual inspection suggests fair agreement within $\sim10-20\%$. The level of agreement among the various studies is encouraging, though somewhat surprising given the differing analyses. Given our thorough methodology (kinematic modelling and beam smearing corrections are more robustly determined, NIR JWST data are considered for SED fitting and surface brightness profiles, and realistic AD corrections are incorporated), we expect our best-fitting parameters to be as accurate as possible, considering the quality of the currently available data.

\begin{table}[h]
\caption{Best-fitting parameters ($\alpha$, $\beta$, $\epsilon_\perp$) and vertical RMS scatter ($\sigma_{M_\ast}$ for the TFR and $\sigma_{j_\ast}$ for the FR) of our $z=0.9 $ TFR and FR.} \vspace{-0.1cm}
\begin{center} 
\resizebox{0.5\textwidth}{!}{
\begin{tabular}{lccccc}
	\hline \noalign{\smallskip}
& Scaling law & $a$ & $b$ & $\epsilon_\perp$ & $\sigma_{M_\ast}$ or $\sigma_{j_\ast}$  \\ \noalign{\smallskip} \hline \noalign{\smallskip}
  & $M_\ast-V_{\rm circ,f}$ & $3.82^{+0.55}_{-0.40}$ & $10.27^{+0.06}_{-0.07}$ & $0.06^{+0.01}_{-0.01}$ & $0.26$\\ \noalign{\smallskip}
  &  $j_\ast-M_\ast$ & $0.44^{+0.06}_{-0.06}$ & $2.86^{+0.02}_{-0.02}$ & $0.11^{+0.02}_{-0.01}$ & $0.13$\\ \noalign{\smallskip}
    \hline 
\end{tabular}} \vspace{-0.2cm}
  \end{center}
  \label{tab:coeff_z1}
\end{table}

\subsection{Evolution since the Universe's half-age}
\label{sec:evolution}

Here we explore whether the scaling relations have evolved since $z\sim1$. Among the many low-$z$ measurements, we adopt \cite{marasco_mstar} as our $z=0$ reference for three main reasons: 1) their analysis is based on the SPARC sample \citep{sparc}, which we used to calibrate our methods (Appendix~\ref{app:sparc}); 2) their $M_\ast$ estimates are based on the same technique as ours; and 3) they adopt the same pivot values in their fitting ($10^{10.5}\,M_\odot$ and $150\ \rm{km/s}$), which simplifies the comparison. We note that \cite{marasco_mstar} used directly H\,{\sc{i}} kinematics to estimate $V_{\rm circ,f}$ and $j_\ast$ without applying AD corrections, but we do not expect this to introduce big differences since rotation-to-dispersion ratios of H\,{\sc{i}} data in the nearby Universe are much larger than those of our sample, leading to small differences between $V_{\mathrm{H\,\textsc{i}}}$, $V_{\rm circ,f}$, and $V_{\ast}$, especially at $M_\ast>10^8\,M_\odot$ \citep{paperIBFR}. \cite{marasco_mstar} reports a slope $a=5.21\pm 0.18$, intercept $b = 10.16\pm0.03$, intrinsic scatter $\epsilon_\perp = 0.039$, and vertical observed scatter $\sigma_{M_\ast}=0.274$ for the local TFR, and $a=0.52\pm 0.03$, $b = 3.04\pm0.03$, $\epsilon_\perp = 0.128$, and $\sigma_{j_\ast}=0.171$ for the local FR (similar to e.g. \citealt{postijstar,paperIBFR}). 

\begin{figure}
    \includegraphics[width=1\linewidth]{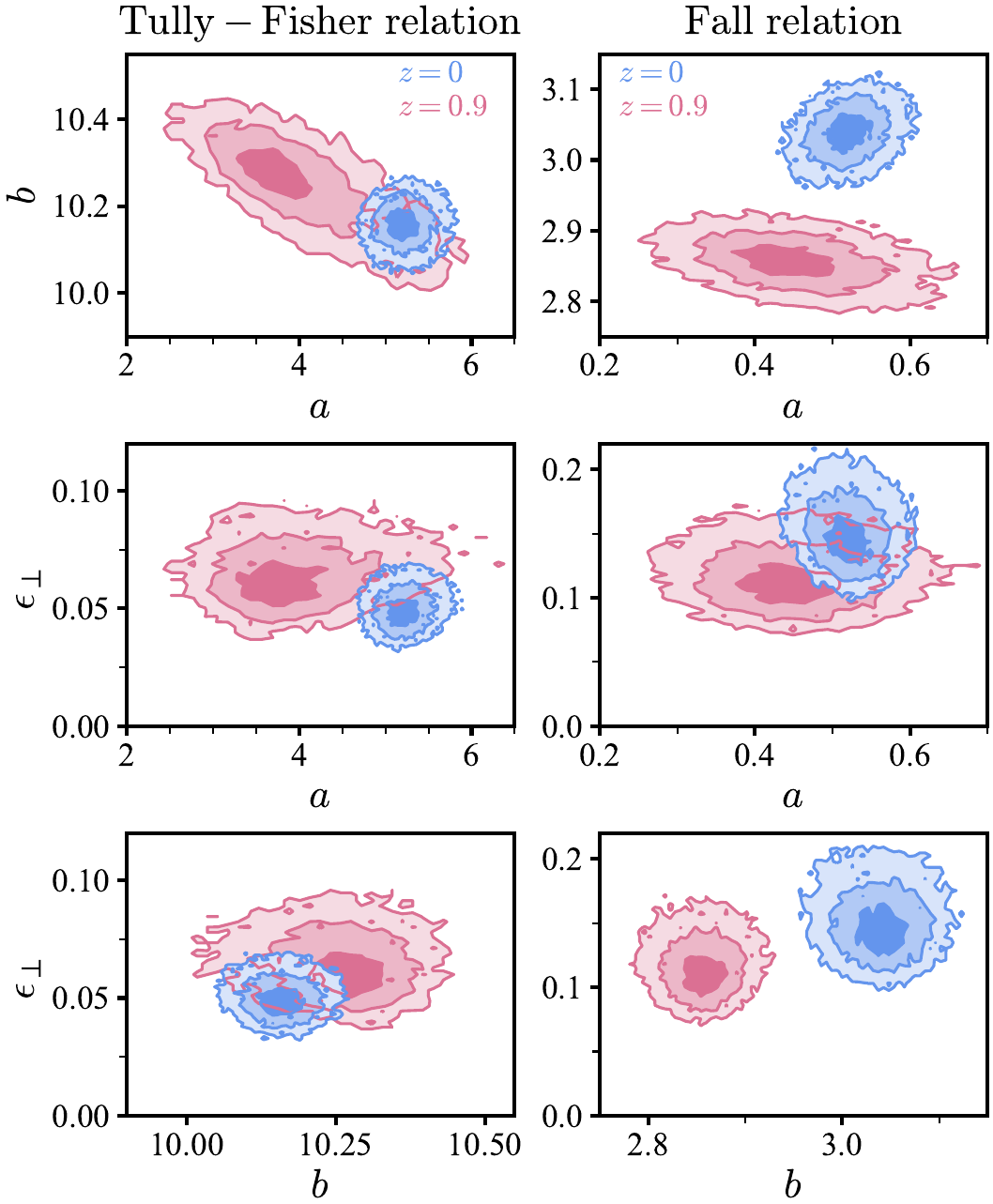}
    \caption{Posterior distributions of the best-fitting TFR and FR. The distributions of low- and high-$z$ disc galaxies are shown in blue (from \citealt{marasco_mstar}) and pink (this work), respectively.
    The contours encompass the 0.393, 0.865, and 0.989 percentiles, corresponding to 1, 2, and $3\,\sigma$ in 2D distributions.}
    \label{fig:evolution2D} \vspace{-0.3cm}
\end{figure}

As shown in Fig.~\ref{fig:relationsz1}, our TFR and FR at $z=0.9$ differ from those at $z=0$. The evolution of the relations can also be visualised in Fig.~\ref{fig:evolution2D}, where we compare the posterior distributions of our best-fitting parameters to those of \cite{marasco_mstar}. As seen from Fig.~\ref{fig:evolution2D} (note that unaccounted uncertainties in masses and kinematics may further broaden the confidence levels), our results present moderate evidence of the evolution of the TFR, and strong evidence of the evolution of the FR. In both relations, the evolution is driven by changes in slopes and intercepts, rather than by the intrinsic scatter. Near $z\sim1$, the TFR is shallower and has a higher intercept by about 0.1 dex. On the other hand, the FR has a slightly shallower slope (unlike the substantial changes reported by \citealt{espejo2025} at $z\approx1.5-2.5$) and its intercept is lower by $\sim0.2$ dex. This latter offset has also been found by theoretical studies based on disc-stability conditions \citep{obreschkow2015} and semi-analytical models \citep{stevens2016}, as well as observational studies \citep{kross_harrison,swinbank2017}.

\subsection{Caveats and systematics}
\label{sec:caveats}

Here, we discuss caveats in our analysis and the potential for systematic errors to affect our results. First, we investigate the hypothetical case in which our observational uncertainties may have been somewhat underestimated (e.g. by our use of $z=0$ calibrations to estimate $\sigma_{z,\ast}$, or the assumption of constant scale heights in Sect.~\ref{sec:method}; note that otherwise our uncertainties account for the various individual uncertainties that enter our derivation of $V_{\rm circ,f}$ and $j_\ast$). Since underestimated errors can bias the recovery of the relations \citep{ubler2017,alcorn2018,espejo2025}, we aim to estimate a realistic additional error budget for our observations to assess the robustness of our best-fitting parameters. A reasonable approach is to inflate the uncertainties in $\log(V_{\rm circ,f})$ and $\log(j_\ast)$ by an amount so that the intrinsic perpendicular scatter ($\epsilon_\perp$) in both scaling relations is nullified. This turns out to be $0.05$ dex for the former and $0.1$ dex for the latter. However, refitting the TFR and FR shows that such inflation has no repercussion on our results, since the best-fitting slopes and intercepts remain unchanged. Following \cite{ubler2017}, we also explored by how much our uncertainties in $\log(M_\ast/M_\odot)$ would need to be underestimated to make the $z=0.9$ TFR consistent with the local one from \cite{marasco_mstar}, finding that only a severe underestimation by $0.4$ dex would do it, although this would also result in a stronger evolution of the FR (decreasing its slope by 0.1).

Next, we investigate two additional questions: Could there be systematic effects that affect the observed evolution in our TFR and FR? Could our sample be missing a galaxy population that, when included, would weaken the evolution of the relations? We start with the first question. Considering the direction of the evolution in the TFR, we are particularly interested in a possible systematic error or bias that results in $V_{\rm circ,f}$ measurements progressively biased low for decreasing $M_\ast$, since this could cause the shallower slope and higher intercept in the $z=0.9$ TFR compared to the $z=0$ TFR. We find it unlikely that this is the case. First, we note that the typical effect of the AD correction is to increase the velocities of low-mass galaxies; therefore, ignoring it would amplify the difference rather than reduce it. In fact, without AD corrections, the TFR and FR would have changes of slopes and intercepts given by $\delta a=a-a_{\rm noADC}=0.5$ and $\delta b=b-b_{\rm noADC}=-0.17$ dex for the TFR, and $\delta a=-0.05$ and $\delta b=0.02$ dex for the FR. This highlights the importance of applying the AD corrections described in Sect.~\ref{sec:method}, which we encourage future studies to incorporate.

Next, we can examine if any factors within our calculation of $V_{\rm AD,\ast}$ could lead to underestimating $V_{\rm circ,f}$ for decreasing $M_\ast$. For this, we would need to have systematically underestimated $\sigma_{\mathrm{H}\alpha}$ and/or overestimated $R_{\mathrm{eff,H}\alpha}$. A strong underestimation of $\sigma_{\mathrm{H}\alpha}$ is unlikely, as our kinematic software has been shown to recover gas velocity dispersions within $20\%$ for data with spatial resolution and S/N like ours \citep{barolo}, and an underestimation of $20\%$ on $\sigma_{\mathrm{H}\alpha}$ at our lowest rotational speeds ($\sim120\,\rm{km/s}$) leads to a $V_{\rm circ,f}$ higher by $10\%$ at most (0.04 dex in $\log(V_{\rm circ,f})$). Our assumed $R_{\mathrm{eff,H}\alpha}=1.13(\pm0.05)\,R_{\mathrm{eff},\ast}$ is well calibrated at $z=1$ \citep{nelson2016,wilman2020}, and since $R_{\mathrm{eff,H}\alpha}>R_{\mathrm{eff},\ast}$ is a condition for the well known inside-out growth of galaxy discs \citep{nelson2012,nelson2016}, there is little room for an overestimation. Fixing $R_{\mathrm{eff,H}\alpha}=R_{\mathrm{eff},\ast}$ plus a $20\%$ underestimation of $\sigma_{\mathrm{H}\alpha}$ would only lead to an increase in $\log(V_{\rm circ,f})$ of around 0.03, 0.02, 0.01, and 0.005 dex at $\log(M_\ast/M_\odot)=10, 10.5, 11, 11.5$, which is largely insufficient to account for the evolution shown by the TFR (see also Sect.~\ref{sec:evolution}). 

Two additional scenarios that could bias our TFR slope towards shallower values are if we systematically overestimate $M_\ast$ for low-luminosity galaxies and/or if we systematically overestimate disc inclinations at low $M_\ast$. We also deem these cases unlikely. On the one hand, our $M_\ast$ recovery methods have been calibrated at $z=0$, yielding excellent agreement with dynamical estimates and stellar population models \citep{marasco_mstar}. On the other hand, our inclinations are derived from accurate modelling of JWST and HST imaging of our galaxies (which we find to be in good agreement with H$\alpha$ morphology inclinations) and, most importantly, show no correlation with the location of galaxies in the TFR and FR. We also note that in Sect.~\ref{sec:data} we assumed the same intrinsic thickness $q_0=0.2$ for all galaxies when converting the observed axis ratios to inclinations, but if we had assumed thicker discs for lower-mass galaxies, their inclinations would be larger, leading to lower rotational speed, hence to a shallower TFR slope. So far, we have focused on our $z=0.9$ TFR, but it is crucial to note that systematic corrections increasing the velocities at the low $M_\ast$ end more than at the high $M_\ast$ end would drive a more substantial evolution of the FR at $z=0.9$, as it would become shallower.\smallskip

We turn now to the second question formulated above, that is, whether we could be missing a subset of the star-forming galaxy population that would change our TFR and FR and weaken their evolution. Given our imposed selection criteria, which require only resolved galaxies with clear velocity gradients, we may be missing the most compact and slowly rotating galaxies within our observed $M_\ast$ range. However, by comparing our sample against the $z=0.9$ stellar mass-size relation from \cite{martorano2024} in Fig.~\ref{fig:sample}, it becomes evident that although the scatter of our sample is not as wide as the full distribution from \cite{martorano2024} and we may lack the rare smallest systems, we should not be missing a significant fraction of compact galaxies at any $M_\ast$. Nevertheless, note that, due to surface brightness dimming, the $z\sim1$ samples may undersample low-surface-brightness galaxies compared to the $z=0$ sample. Here, it is key to note that if we were missing slowly rotating or very compact galaxies at all masses, the intercepts of our TFR and FR would increase and decrease, respectively, driving a more substantial evolution than reported above. To weaken the evolution in our relations, we would need to be missing a significant population of fast-rotating galaxies at all masses below $\log(M_\ast/M_\odot)\lesssim10.8$, which does not seem feasible, as there is no reason to expect such a population would go undetected.

Finally, we also acknowledge that if the TFR and FR at $z=0$ and $z=0.9$ have some curvature, the more limited $M_\ast$ span of our sample (reaching $\log(M_\ast/M_\odot)>9.5$) compared to the $z=0$ samples (reaching $\log(M_\ast/M_\odot)\sim7$, e.g. \citealt{marasco_mstar,paper_galaxyhalo}) could affect the best-fitting power-law parameters. For example, fitting the $z=0$ data from \cite{marasco_mstar} but masking out all galaxies with $\log(M_\ast/M_\odot)<9.5$ results in different local relations. On the one hand, the TFR with the mass cut has a slope that is 0.74 lower and an intercept that is 0.09 dex higher. On the other hand, the FR cut with the mass cut has a slope that is 0.13 higher and an intercept that is 0.04 dex lower. Understanding precisely the actual effect on our $z=0.9$ relations will only be possible with the next generation of IFUs to be installed at the ELT, and partially with upcoming ERIS data, which has the same spatial resolution as KMOS but better spectral resolution.

In summary, the potential systematic effects discussed above appear too small to significantly affect our TFR and FR (particularly if they are unbroken power laws). Unless additional, as yet unidentified, systematics are at play, our analysis suggests evolution in both relations.

\section{Links with dark matter haloes' scaling relations}

\subsection{Drivers of the evolution}
\label{sec:origin}
\begin{figure*}
\includegraphics[width=1\linewidth]{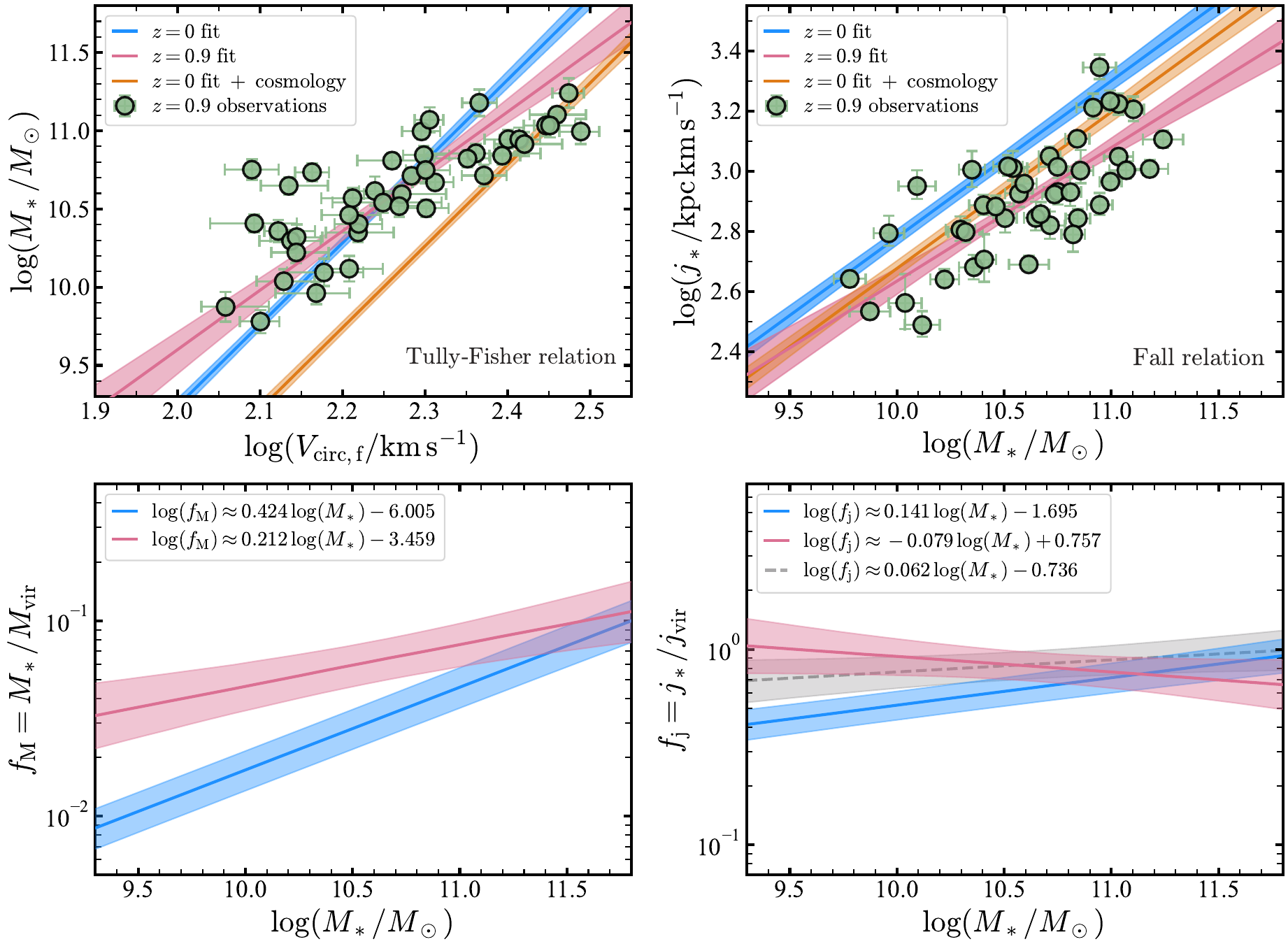}
    \caption{Drivers of the evolution in our scaling relations. The \emph{top} panels compare how the local TFR and FR (blue curves) would evolve (orange curves) if the only changes are the $z$ evolution of the Hubble parameter and the density contrast, against our high-$z$ data (green markers) and best fits (pink curves). The results imply an evolution of $f_{\rm V}f_{\rm M}^{-1/3}$ and $f_{\rm j}f_{\rm M}^{-2/3}$.
    The \emph{bottom} panels show $f_{\rm M}$ and $f_{\rm j}$ at $z=0$ (blue) and $z=0.9$ (pink) as a function of $M_\ast$, as implied by our best-fitting relations after assuming $f_{\rm V}=1.3\pm0.1$. 
    The dashed grey line and bands in the bottom right panel show $f_{\rm j}$ at $z=0.9$ under the assumption that $f_{\rm M}$ does not evolve with redshift.
    In all the panels, bands correspond to $1\,\sigma$ confidence bands around the mean relations. Note that the vertical axis span in the bottom panels is the same, highlighting the stronger mass dependency of $f_{\rm M}$ over $f_{\rm j}$.}
    \label{fig:fractions}
\end{figure*}

Next, we examine some simple consequences of our empirical findings in the context of the link between the scaling relations of disc galaxies and their dark matter haloes. We start with the virial relations for dark matter haloes in CDM-like cosmologies (see \citealt{bookFilippo}):
\begin{equation}
M_{\rm vir} = \frac{1}{G\, H(z)} \sqrt{\frac{2}{\Delta_{\rm c}(z)}}\, V_{\rm vir}^3~,
\end{equation} \vspace{-0.3cm}
\begin{equation}
\label{eq:jvir}
j_{\rm vir} = \frac{2\, \lambda}{H(z)\, \sqrt{\Delta_{\rm c}(z)}}\, V_{\rm vir}^2~,\end{equation}
where $G$ is the gravitational constant, $\Delta_{\rm c}(z)$ the critical density for virialisation, $H(z)$ the Hubble parameter, and $\lambda$ the \cite{bullock2001_am} spin parameter which follows a log-normal distribution peaked at $\log\lambda\approx-1.456$, irrespective of $z$ and $M_{\rm vir}$ (but see also \citealt{bett2007}, \citealt{teklu_angularmomentum}, and \citealt{rodriguezgomez2017} for potential subtle dependencies on $M_{\rm vir}$, $z$, and morphology). These equations\footnote{Eq.~\ref{eq:jvir} is exact for haloes with singular isothermal sphere (SIS) profiles, and accurate within $\sim10\%$ for NFW haloes \citep{nfw} with $10 < \log(M_{\rm vir}/M_\odot) < 14$ (e.g. \citealt{bullock2001_am}).}  can be rewritten in terms of the observables $M_\ast$, $V_{\rm circ,f}$, and $j_\ast$, and the ratios $f_{\rm V}=V_{\rm circ,f}/V_{\rm vir}$, $f_{\rm j}=j_\ast/j_{\rm vir}$, and $f_{\rm M}=M_\ast/M_{\rm vir}$ such that
\begin{equation}
\label{eq:vf}
    M_\ast = \frac{\sqrt{2}\, \left( f_{\rm M} / f_{\rm V}^3 \right)}{G\, H \, \sqrt{\Delta_{\rm c}}}\, V_{\rm circ,f}^3~,
\end{equation}
\begin{equation}
\label{eq:js}
 j_\ast = \frac{2\, \lambda \left( f_{\rm j} / f_{\rm V}^2 \right)}{H\, \sqrt{\Delta_{\rm c}}} \, V_{\rm circ,f}^2~.
\end{equation}

\noindent Evaluating these equations at redshift $z$ and $z=0$, we have
\begin{equation}
\label{eq:v_evol}
\dfrac{V_{\rm circ,f}(M_\ast,z)}{V_{\rm circ,f}(M_\ast,0)} = 
\left[ \dfrac{f_{\rm V}(M_\ast,z)}{f_{\rm V}(M_\ast,0)}   \right]
\left[ \dfrac{f_{\rm M}(M_\ast,z)}{f_{\rm M}(M_\ast,0)}   \right]^{-1/3}
\left[ \dfrac{\Delta_{\rm c}(z)}{\Delta_{\rm c}(0)}  \right]^{1/6}\,
\left[ \dfrac{H(z)}{H(0)}  \right]^{1/3}~,
\end{equation}
\vspace{-0.2cm}
\begin{equation}
\label{eq:j_evol}
\frac{j_\ast(M_\ast, z)}{j_\ast(M_\ast, 0)} =
\left[ \frac{f_j(M_\ast, z)}{f_j(M_\ast, 0)} \right]
\left[ \frac{f_{\rm M}(M_\ast, z)}{f_{\rm M}(M_\ast, 0)} \right]^{-2/3}
\left[ \frac{\Delta_{\rm c}(z)}{\Delta_{\rm c}(0)} \right]^{-1/6}
\left[ \frac{H(z)}{H(0)} \right]^{-1/3},
\end{equation}
\noindent which give the $z$ evolution of the TFR and FR.

We first test the hypothetical scenario in which the TFR and FR evolve exclusively through the redshift dependence of the $H$ and $\Delta_{\rm c}$. In this case, $f_{\rm M}$, $f_{\rm j}$, and $f_{\rm V}$ are assumed to be independent of $z$ (although may depend on $M_\ast$), and the first two factors on the right hand side of Eqs.~\ref{eq:v_evol} and \ref{eq:j_evol} are unity. With these expressions under the cosmology-only halo scaling hypothesis, we can propagate the TFR and FR from \cite{marasco_mstar} back to $z=0.9$ and compare with our observed $z=0.9$ relations. For our cosmology, we have $H(z=0.9)=116.3\,\rm{km\,s^{-1}\,Mpc^{-1}}$, $\Delta_{\rm c}(z=0)=101$ and $\Delta_{\rm c}(z=0.9)=154$ \citep{bryan1998}. The results of this exercise are shown in the top panels of Fig.~\ref{fig:fractions}. The evolved TFR and FR are even more discrepant with respect to the observed $z=0.9$ relations than those observed at $z=0$. Note also that the systematic offsets needed to match our data with the cosmology-only halo scaling hypothesis are significantly larger than those discussed in Sect.~\ref{sec:caveats}. Therefore, to match the data, the factors $f_{\rm V}f_{\rm M}^{-1/3}$ and $f_{\rm j}f_{\rm M}^{-2/3}$ must evolve.\smallskip

At this stage, it is instructive to rearrange Eqs.~\ref{eq:vf} and \ref{eq:js} as
\begin{equation}
\label{eq:fm}
    f_{\rm M}(M_\ast,z) =\sqrt{\dfrac{\Delta_{\rm c}(z)}{2}}\,G\,H(z)\,M_\ast\, \left(\dfrac{f_{\rm V}(M_\ast,z)}{V_{\rm circ,f}(M_\ast,z)}  \right)^3 ,\ \rm{and}
\end{equation}
\begin{equation}
\label{eq:fj}
f_{\rm j}(M_\ast,z) = 
\frac{H(z)\sqrt{\Delta_{\rm c}(z)}}{2\,\lambda} 
j_\ast(M_\ast,z) 
\left( \frac{f_{\rm V}(M_\ast,z)}{V_{\rm circ,f}(M_\ast,z)} \right)^2 ~.
\end{equation}

\noindent
From these equations, it is clear that $f_{\rm M}$ and $f_{\rm j}$ at any $z$ are entirely specified by the cosmological redshift dependence of the Hubble parameter and density contrast, by $f_{\rm V}$, and by the observed TFR and FR. The ratio $f_{\rm V}$ depends on the baryonic and dark matter expansion and contraction, but varies only weakly with $M_\ast$ (see \citealt{dutton2010,mcgaugh_mondBTFR}). For instance, \cite{reyes2012} combined kinematic and weak-lensing observations and found $f_{\rm V}\approx1.3-1.4$ for galaxies with $9.8<\log(M_\ast/M_\odot)<10.8$ at $z=0$. In addition, results from \cite{postinomissing} and \cite{paper_galaxyhalo} on mass models from rotation curve decomposition imply in $f_{\rm V} \approx1.2-1.4$ for galaxies with $100 < V_{\rm circ,f}/\rm{km\ s^{-1}} < 350$. Based on this, and for simplicity, we neglect any $M_\ast$ dependency and consider $f_{\rm V}=1.3$ with a $1\,\sigma$ scatter of 0.1. We further assume that $f_{\rm V}$ does not change with $z$; although no observational constraints exist, idealised models suggest that $f_{\rm V}$ evolves significantly only at $z\gtrsim2$ \citep{somerville2008,dutton2011}.

From the best-fit TFR and FR form this work and \cite{marasco_mstar}, we compute the resulting ratios, shown in the bottom panels of Fig.~\ref{fig:fractions}. Solid lines show the median relations, and the confidence bands the 16th-84th percentile range, derived from the sampling of the best-fitting TFR and FR (including their intrinsic scatter $\epsilon_\perp$). For the $f_{\rm M}-M_\ast$ and $f_{\rm j}-M_\ast$ relations at $z=0$, the width is around 0.08--0.1 dex. For the $z=0.9$ relations, the width is around 0.10--0.13 dex, respectively. For completeness, we also carry out the exercise of propagating the observed vertical scatter in the TFR ($\sigma_{M_\ast}$) and FR ($\sigma_{j_\ast}$) into the $f_{\rm M}-M_\ast$ and $f_{\rm j}-M_\ast$ relations. At $z=0$, $\sigma_{M_\ast}$ ($\sigma_{j_\ast}$) propagates into a scatter of 0.16 (0.20) dex at fixed $M_\ast$ (slightly smaller but comparable to literature values, e.g. \citealt{romeo2020_instabilities,romeo2023}). At $z=0.9$, $\sigma_{M_\ast}$ ($\sigma_{j_\ast}$) propagates into a scatter of 0.20 (0.19) dex at fixed $M_\ast$. Overall, our redshift comparison reveals changes in $f_{\rm M}$ and $f_{\rm j}$ from $z=0.9$ to the present, which we discuss next.

\subsection{The evolution of $f_{\rm M}$}
\label{sec:fm}
The bottom left panel of Fig.~\ref{fig:fractions} shows $f_{\rm M}$ at $z=0$, exhibiting the same single power-law behaviour found for nearby star-forming galaxies based on rotation curve decomposition (e.g. \citealt{postinomissing,enrico_massmodels_ss,paper_galaxyhalo}). In particular, $f_{\rm M}$ at $z=0$ is well described by the expression $\log(f_{\rm M}) = 0.424\log(M_\ast/M_\odot) -6.002$. On the other hand, at $z=0.9$ we find higher $f_{\rm M}$ values for $M_\ast<10^{12}\,M_\odot$ and, crucially, a weaker $M_\ast$ dependency driven by the shallower TFR at $z=0.9$, with $\log(f_{\rm M}) = 0.212\log(M_\ast/M_\odot) -3.459$. The slope of the $\log(f_{\rm M})-\log(M_\ast)$ relations at $z=0.9$ and $z=0$ differ at the same level as the slopes of the TFRs, i.e. around $2\,\sigma$.

\begin{figure}
    \centering
    \includegraphics[width=1\linewidth]{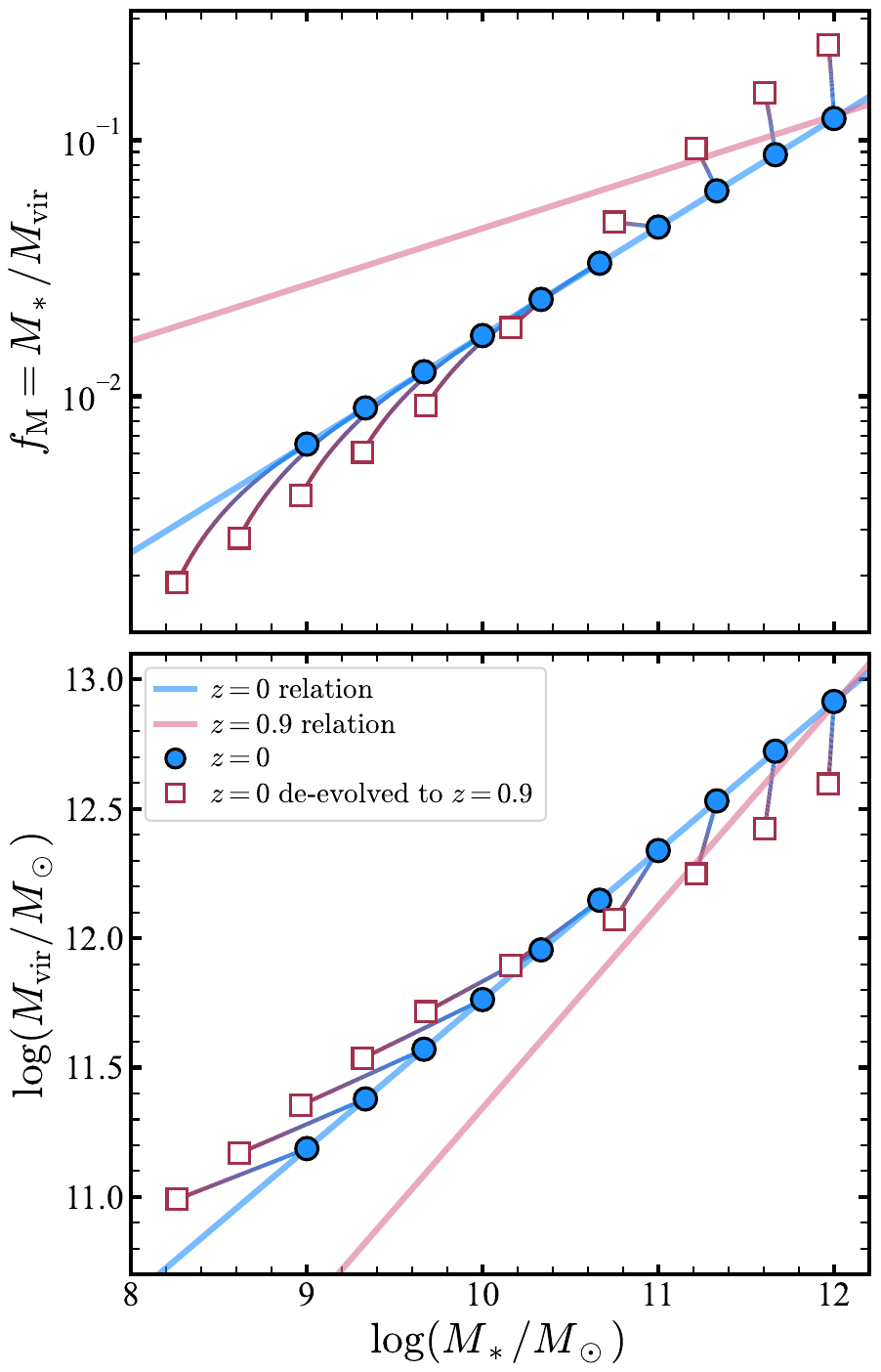}
     \vspace{-0.4cm}
     \caption{Comparison between our inferred $f_{\rm M}-M_\ast$ (\emph{top}) and $M_{\rm vir}-M_\ast$ (\emph{bottom}) relations and an idealised toy model of mass assembly. The model considers a population of galaxies (blue circles) lying in the $z=0$ relations (blue curves), which are then traced back to $z=0.9$ (squares), assuming theoretical stellar and halo mass growths. Galaxy downsizing and halo growth histories yield $z=0.9$ relations (squares) with slopes different from those observed at $z=0$ (solid pink curves). 
     The only way to reconcile the stellar mass growth of galaxies with our inferred relations at $z=0.9$ is if low-mass haloes grew more than high-mass haloes since $z=0.9$, contrary to CDM expectations. This suggests that the galaxies populating our $z=0.9$ relations do not follow this toy model and are unlikely to be the progenitors of the $z=0$ population.} \vspace{-0.3cm}
    \label{fig:sketch}
\end{figure}

Our observed scaling relations can be used to set interesting constraints in the broader context of galaxy assembly. In particular, we now show that pairing the CDM halo assembly framework with our observational results implies that
main-sequence $z=0.9$ disc galaxies cannot evolve into the local main-sequence disc population following a simple evolutionary model in which they remain in the SFMS from $z=0.9$ to $z=0$. To show this schematically, in the top panel of Fig.~\ref{fig:sketch} we construct a simple toy model in which the local $f_{\rm M}-M_\ast$ relation (blue curve) is populated with mock galaxies (circles) and evolved backwards in time to $z=0.9$ (squares). The bottom panel recasts the same comparison into the more familiar stellar-to-halo mass plane, highlighting the relative individual growth of $M_\ast$ and $M_{\rm vir}$. For this exercise, we assume that the galaxies' stellar mass evolves following the main sequence $\mathrm{star\ formation\ rate}-M_\ast-z$ relation of \citet{leja2022}, and that their halo growth follows the halo accretion formalism by 
\citet{vandenbosch2014}. A toy model such as this entails mappings between $M_\ast(z=0.9)$ and $M_\ast(z=0)$ and between $M_{\rm vir}(z=0.9)$ and $M_{\rm vir}(z=0)$ that are monotonic, one-to-one, and rank-order preserving. As a result, the model neglects many complexities of galaxy and halo evolution, as discussed below.

The mass growths implied by such a model yield relations at $z=0.9$, which depart from our measurements (pink curves). In general, the bottom panel of Fig.~\ref{fig:sketch} shows that any plausible stellar-mass assembly track would imply a paradoxical, anti-hierarchical growth of DM halos, which would undermine our entire interpretive framework to begin with. This becomes evident when one draws arbitrary evolutionary tracks in $M_\ast$ between the $z=0.9$ and $z=0$ relations in the bottom panel of Fig.~\ref{fig:sketch}, under the minimal assumption that masses do not decrease with time. Independent of the specific increment in $M_\ast$, massive haloes must already be largely assembled by $z=0.9$, while low-mass systems would still be building a substantial fraction of their halo mass at later times\footnote{We note that sensible variations in $f_{\rm V}$ cannot explain the discrepancy. For example, reducing $f_{\rm V}$ by $10-20\%$ (e.g. \citealt{dutton2011}) leaves the slopes unchanged and does not shift the $z=0.9$ relations relative to those at $z=0$.
Matching the data requires an implausibly strong mass dependence of $f_{\rm V}$ \citep{postinomissing,paper_galaxyhalo} and $f_{\rm V}<1$ for $V_{\rm circ,f}<250\,\mathrm{km\,s^{-1}}$, which is unphysical \citep{mcgaugh_mondBTFR}.}. Such behaviour is at odds with the well-established theoretical expectations in the CDM paradigm, where massive haloes experience stronger fractional growth at late times than their lower-mass counterparts \citep[e.g.][]{fakhouri2010, vandenbosch2014, correa2015}.

Therefore, we conclude that, if our scaling relations are correct, our main-sequence disc $z=0.9$ sample cannot follow the simplified toy model described above, and it is likely not the progenitor of our local main-sequence disc sample (i.e. we are witnessing an example of potential progenitor bias; e.g., \citealt{vandokkum1996,vandokkum2001}). This is perhaps not surprising, as the idealised toy model above does not capture processes such as mergers, feedback, star-formation bursts, and quenching, all of which can lead to morphological transformations \citep{bookFilippo}, which our results suggest are in place.

As another option, we can consider the possibility that our observed scaling relations are incorrect. As shown in the top-left panel of Fig.~\ref{fig:fractions}, reconciling the $z=0.9$ TFR with the cosmology-only halo scaling hypothesis (i.e. assuming no evolution of $f_{\rm M}$) would mean that the circular speeds of galaxies with $\log(M_\ast/M_\odot)\lesssim10.7$ have been underestimated by $\sim50\%$. As discussed extensively in Sect.~\ref{sec:caveats}, the uncertainties considered in this work appear to be insufficient to account for this discrepancy on their own. Another possibility is that strong selection-function effects, additional unknown systematic effects, or inappropriate assumptions may bias our results. 
Crucially, we note that such effects must have also been present in previous TFR determinations at $z\sim1$, which report slopes consistent with ours (see Sect.~\ref{sec:scaling}). Alternatively, one may be inclined to conclude that ionised gas kinematics cannot be used to infer dynamical properties of high-$z$ star-forming galaxies, regardless of a careful approach such as ours, which includes robust 3D kinematic modelling, AD corrections, Sérsic modelling of NIR data, and homogeneous stellar mass determinations. On the theoretical side, this should be further explored through simulations that track the evolution of individual galaxies. On the observational side, it would be key to trace the TFR and FR across finer redshift intervals and to earlier epochs. Upcoming facilities such as SKA, ALMA, Euclid, and Roman will also provide direct constraints on galaxy kinematics and halo properties through deep cold-gas observations and weak-lensing measurements, and will be key to obtaining a more definitive picture.

Finally, we would like to highlight the recent studies based on abundance matching and angular clustering by \cite{shuntov2025} and \cite{paquereau2025}, respectively. Looking at the COSMOS field with JWST, the above authors also find tentative evidence of higher values of $f_{\rm M}$ at $z\sim1$ than at $z\sim0$, though no explanation has yet been offered. This is unlike previous abundance-matching studies\footnote{In Appendix~\ref{app:AM}, we show the TFR required to match the $f_{\rm M}$ from abundance matching from \cite{moster2010}. Note that abundance matching also appears to be in tension with $f_{\rm M}$ measurements at $z=0$ derived from rotation curves \citep{postinomissing,enrico_massmodels_ss,paper_galaxyhalo}.}, which typically suggest a higher $f_{\rm M}$ at $z=0$ than at $z=1$, with only mild evolution in slope \citep{moster2010}. State-of-the-art hydrodynamical simulations generally show little redshift evolution in $f_{\rm M}$ \citep{colibre_shmr}, so clearly obtaining additional observational constraints will be key.

\subsection{The evolution of $f_{\rm j}$ and $f_{\rm R}$}
Next, we turn our attention to $f_{\rm j}$, shown in the bottom right panel of Fig.~\ref{fig:fractions}. At both redshifts, $f_{\rm j}$ depends more weakly on $M_\ast$ than $f_{\rm M}$ (see also e.g. \citealt{posti_galaxyhalo,enrico_massmodels_ss,romeo2023}). Within our mass coverage, $f_{\rm j}\sim 0.8$, in agreement with previous studies (e.g. \citealt{fall2013, fall2018,posti_galaxyhalo,enrico_massmodels_ss}). Specifically, within $9.5 < \log(M_\ast/M_\odot)<11.5$, $f_{\rm M}$ at $z=0$ increases by a factor of $\sim2$, while at $z=0.9$ it decreases by $30\%$. The $f_{\rm j}-M_\ast$ relations are well described by the power laws $\log(f_{\rm j}) = 0.141\log(M_\ast/M_\odot) -1.695$ at $z=0$, and $\log(f_{\rm j}) = -0.079\log(M_\ast/M_\odot) +0.757$ at $z=0.9$. 

Taken at face value, our results indicate that disc galaxies with $\log(M_\ast/M_\odot)\lesssim 11.1$ have lowered their $f_{\rm j}$ over the last $\sim8$~Gyr and more massive galaxies have raised it, but the overall redshift and mass dependence is weak. Since $f_{\rm j}$ depends indirectly on $f_{\rm M}$ through the TFR ($f_{\rm j}\propto f_{\rm M}^{2/3}$; see Eq.~\ref{eq:fj}), it is instructive to explore a limiting case in which $f_{\rm M}$ does not evolve. In this scenario, the evolution of $f_{\rm j}$ arises exclusively from the Hubble parameter, the density contrast, and the FR. This assumption is not fully self-consistent, as it requires that our measurements trace $V_\ast$ but not $V_{\rm circ,f}$. Nevertheless, it is helpful as a hypothetical exercise to test whether an evolution of $f_{\rm j}$ remains. The results are shown in the bottom-right panel of Fig.~\ref{fig:fractions} as a dashed grey curve. The imposed steeper $f_{\rm M}$-$M_\ast$ relation flips the sign of the $f_{\rm j}$-$M_\ast$ slope, though the dependence on $M_\ast$ remains weak. As already suggested by the orange curve in the top-right panel of Fig.~\ref{fig:fractions} (if $f_{\rm M}$ and $f_{\rm j}$ do not evolve with $z$ the observed FR is not reproduced), the conservative scenario of no evolution in $f_{\rm M}$ still results in an evolution of $f_{\rm j}$, though it is relatively mild.

Different mechanisms can increase or decrease $f_{\rm j}$ as a function of $M_\ast$ and $z$, such as outflows, galactic fountains, stripping, mergers, and dynamical friction (e.g. \citealt{vandenbosch+01,brook2012_angularmomentum,romanowsky,lagos_eagle,irodotou2019}). For instance, feedback-driven outflows near galaxy centres can lead to a larger $f_{\rm j}$ \citep{brook2011,romanowsky}. In addition, the cumulative effect of dry mergers between $z=0.9$ to $z=0$ could reduce $f_{\rm j}$ at $z=0$ relative to $z=0.9$ \citep{fall1979,romanowsky,lagos18}. A combination of these different phenomena could explain the behaviour of $f_{\rm j}$ in the bottom right panel of Fig.~\ref{fig:fractions}. Yet, establishing the precise mechanisms would require a more accurate determination of $f_{\rm j}$, coupled with analysis of hydrodynamical simulations. Overall, the lack of a strong $z$ or $M_\ast$ dependence of $f_{\rm j}$ suggests that the above processes are gentle (or compensated each other) from $z=0.9$ to $z=0$. \\

\noindent
Our results also have immediate implications for the relative sizes of galaxy discs compared to their host haloes. For exponential discs (a good approximation for our sample; see Fig.~\ref{fig:sample}) embedded in SIS haloes, one obtains $f_{R} = R_{\rm eff,\ast}/R_{\rm vir} = (1.678/\sqrt{2})\,\lambda\,f_{\rm j}/f_{\rm V}$ (e.g. \citealt{fall1980,fall83}; see also \citealt{mo98} for the NFW case). For $f_{\rm V}=1.3$, this reduces to $f_{\rm R}\approx0.032\,f_{\rm j}$, so that the evolution of $f_{R}(M_\ast,z)$ is fully specified by that of $f_{\rm j}(M_\ast,z)$ in Fig.~\ref{fig:fractions}. Our finding of only weak variation in $f_{R}(M_\ast,z)$ agrees well with previous semi-empirical studies (e.g. combining observed galaxy sizes with abundance-matching halo sizes; \citealt{huang2017,somerville2018}) and with results from hydrodynamical simulations \citep{grand2017,rodriguezgomez2022,somerville2025}.

\subsection{A short remark on gas content}

An important aspect to keep in mind is that throughout this study, we have only considered the stellar component of $f_{\rm M}$ and $f_{\rm j}$, since resolved cold gas measurements at $z\sim1$ remain scarce (for indirect estimates see e.g. \citealt{puech2010}), partly due to gaps between ALMA bands. Yet, at $z=0$ it is well established that cold gas carries a substantial fraction of the angular momentum budget and contributes significantly to the baryonic $j$ of galaxies, except in the most massive discs (e.g. \citealt{paperIBFR,paperIIBFR,romeo2023}). Moreover, the baryon-to-halo mass relation is shallower than the stellar one at $z=0$ \citep{romeo2023,paper_galaxyhalo}, and the higher cold gas fractions at earlier epochs \citep{tacconi_review} make this even more relevant at $z\sim1$.

Additionally, cold gas kinematics would help reduce the observational uncertainties in our analysis (see Sect.~\ref{sec:method}), enabling a more robust reassessment of the evolution of the baryon-to-halo mass and angular momentum ratios. Evidently, incorporating the cold gas reservoir is crucial for an integral picture of the evolution of $f_{\rm M}$ and $f_{\rm j}$.

Beyond the cold gas in discs, a complete understanding of the drivers of $f_{\rm M}$ and $f_{\rm j}$ also requires accounting for the circumgalactic medium and large-scale cold gas, both of which can exchange angular momentum with galaxies. Recent studies have explored this using advanced models and simulations (e.g. \citealt{pezzulli2017,afruni2022,wang2022,afruni2023,liu2025,simons2025,wang2025}), but connecting these predictions to observations remains a key challenge for the years to come.

\section{Summary and conclusions}
\label{sec:conclusions}

In this work, we have built the Tully-Fisher ($M_\ast-V_{\rm circ,f}$, TFR) and Fall ($j_\ast-M_\ast$, FR) relations for a sample of disc galaxies at $z=0.9$. To do so, we first compiled a sample of galaxies with public IFU H$\alpha$ kinematic data and space-based (JWST and HST) imaging. We applied different quality cuts, resulting in a high-quality sample of 43 star-forming rotationally supported discs (Fig.~\ref{fig:sample}) with $5\times10^9 < M_\ast/M_\odot < 2\times10^{11}$ and $0.78 < z < 1.03$, encompassing a cosmic time when the Universe was half its current age. The galaxies in our sample have representative star formation rates, Sérsic indices, and effective radii across our $M_\ast$ and $z$ ranges.

\looseness=-1
We derived robust kinematic models (Fig.~\ref{fig:pvs}) that enabled us to retrieve the intrinsic kinematics of the galaxies, despite the low spatial resolution, thereby improving upon previous literature. To account for the pressure-supported motions in our galaxies, we computed realistic asymmetric drift corrections, which allowed us to convert our measured H$\alpha$ rotational velocities into circular speeds (whose flat value $V_{\rm circ,f}$ enters the TFR) and stellar rotational velocities ($V_\ast$, which is needed to compute $j_\ast$ for the FR). 

Paired with spatial and spectral modelling of JWST and HST data, we measured $M_\ast$, $j_\ast$, and $V_{\rm circ,f}$ (in a consistent way as done at $z=0$), and we determined the $z=0.9$ TFR and FR (Fig.~\ref{fig:relationsz1}, see also Fig.~\ref{fig:j_vsigma}). We parametrised the observed trends with power-laws (Fig.~\ref{fig:relationsz1} and Table~\ref{tab:coeff_z1}). We note that the asymmetric drift corrections have a non-negligible impact on the best-fitting TFR and FR, highlighting the importance of applying them.

\looseness=-1
By comparing our best-fitting relations with a recent determination in the nearby Universe based on cold neutral atomic gas, we detected moderate evolution of the TFR and a more substantial evolution of the FR (Fig.~\ref{fig:evolution2D}). At $z=0.9$, the TFR has a shallower mass dependency and a higher intercept. On the other hand, the FR has a slightly shallower slope and an intercept about 0.2 dex lower. We discussed in detail different caveats that could affect the determination of our relations, finding that our results are robust against the presence of potential known systematics (but see below), unless the TFR and FR are not intrinsically unbroken power laws.

A key aspect of this study is that we connected the observed TFR and FR with scaling relations for dark matter haloes. We found that the evolution of the relations cannot be explained by a model where the $z=0$ TFR and FR are evolved to $z=0.9$ by accounting for the $z$ evolution of the Hubble parameter and density contrast alone (Fig.~\ref{fig:fractions}, top panels), but instead requires intrinsic variations in the galaxy mass and angular momentum assembly histories. Specifically, the quantities $f_{\rm V}f_{\rm M}^{-1/3}$ and $f_{\rm j}f_{\rm M}^{-2/3}$ must evolve with $z$, with $f_{\rm V}=V_{\rm circ,f}/V_{\rm vir}$, $f_{\rm M}=M_\ast/M_{\rm vir}$, and $f_{\rm j}=j_\ast/j_{\rm vir}$. 
Choosing a realistic $f_{\rm V}$, we showed the dependencies that $f_{\rm M}$ and $f_{\rm j}$ should have as a function of $M_\ast$ and $z$ given our TFR and FR (Fig.~\ref{fig:fractions}, bottom panels). This represents a significant improvement in the literature studying the $f_{\rm M}$ and $f_{\rm j}$ ratios, which often assume abundance-matching values rather than deriving them directly from the data. At $z=0$, we found that both $f_{\rm M}$ and $f_{\rm j}$ increase with stellar mass, in agreement with previous observations. 

\looseness=-1
Our results show that, within the CDM framework, the observed scaling relations are incompatible with a simple evolutionary model in which our main-sequence disc $z=0.9$ population becomes the progenitor of the main-sequence disc $z=0$ population (Fig.~\ref{fig:sketch}), implying a more complex morphological transformation and baryonic mass assembly instead. Alternatively, it could be that our scaling relations are incorrect due to extreme selection effects, large unrecognised systematics, severely flawed assumptions in our or previous works, or the possibility that ionised-gas kinematics is not a reliable dynamical tracer.

As for $f_{\rm j}$, we found its normalisation to be slightly higher at $z=0.9$ across most of our $M_\ast$ range. However, the overall variation with $M_\ast$ and $z$ is weak, which sets constraints on the different processes that can increase or decrease $f_{\rm j}$ during galaxy formation. The evolution of $f_{\rm j}$ also determines the evolution of the ratio $f_{\rm R }=R_{\rm eff,\ast}/R_{\rm vir}\approx0.032\,f_{\rm j}$. Accordingly, we found relatively little evolution on $f_{\rm R}(M_\ast,z)$, in agreement with previous theoretical and semi-empirical determinations.\\

\noindent
Besides an improvement in the quality of kinematic data, the study of the time evolution of $f_{\rm M}$ and $f_{\rm j}$ needs to be complemented with models and simulations that incorporate the complexity of galaxy evolution in a cosmological context (e.g. \citealt{sales2009,duttonvandenbosch,genel2015,pedrosa2015,teklu_angularmomentum,grand2017,lagos_eagle,elbadry18,rodriguezgomez2022,yang_j_2024}). Our observational results serve as vital benchmarks for such theoretical models.\\

\begin{acknowledgements}
We thank the referee for helpful comments and feedback. We thank Joop Schaye, Filippo Fraternali, Francesca Rizzo, Pieter van Dokkum, and Marijn Franx for valuable comments and discussions. We also thank Christopher Harrison and Mark Swinbank for their clarifications on the KROSS data, Arjen van der Wel for his assistance on Sérsic models, and Henk Hoekstra for clarifications on weak-lensing estimates.

PEMP is funded by the Dutch Research Council (NWO) through the Veni grant VI.Veni.222.364.

EDT was supported by the European Research Council (ERC) under grant agreement no.\ 101040751.

We have used the services from SIMBAD, NED, and ADS extensively, as well as the tool TOPCAT \citep{topcat} and the Python packages NumPy \citep{numpy}, Matplotlib \citep{matplotlib}, SciPy \citep{scipy}, and Astropy \citep{astropy}, for which we are thankful.
\end{acknowledgements}

   \bibliographystyle{aa.bst} 
   \bibliography{references} 

\begin{appendix}
\onecolumn
\section{Kinematic modelling}
\label{app:kinmodels}

Given the limited radial extent of the H$\alpha$ emission in the KROSS and KMOS$^{\rm 3D}$ data relative to the PSF size, the observations can suffer from strong beam smearing \citep{swatersPhD,barolo}. To deal with this, our kinematic approach differs from those conventionally used in other studies of the KROSS and KMOS$^{\rm 3D}$ samples, such as fitting velocity fields or extracting 1D kinematic profiles and applying simplified beam-smearing corrections a posteriori (e.g. \citealt{burkert_j,kross_harrison,swinbank2017}). Such approaches may result in an underestimation of the rotational velocities and an overestimation of the velocity dispersion \citep{barolo,rizzo2022,araujocarvalho2025}.
Instead, we use the software $^{\rm 3D}$Barolo\footnote{v1.7, \url{https://editeodoro.github.io/Bbarolo/}} \citep{barolo}, which performs forward-modelling on the whole H$\alpha$ data cube and, thanks to a convolution with the observational PSF (described by a 2D Gaussian), mitigates the effects of beam-smearing. $^{\rm 3D}$Barolo has been used to fit different emission lines at different redshifts (e.g. \citealt{filippo_highz,sharma2021,enrico_radialmotions,alpaka2023,agc114905_deep,rowland2024,liu2025}), and crucially, it has been tested extensively with artificially degraded H\,{\sc{i}} observations of nearby galaxies and realistic mock galaxies drawn from simulations, showing that it can correctly recover the rotational speed even when the spatial resolution is sparse and reduced down to several kpc (e.g. \citealt{barolo,enrico_z1,huds2020,rizzo2022}).

During our kinematic modelling, some parameters are kept fixed, while others are free to vary. The parameters that remained fixed to our input values are the inclination (fixed to the inclination derived from the \textsc{galfit} fits to the JWST/HST data) and redshift (taken from the KROSS and KMOS$^{\rm 3D}$ databases). Additionally, $^{\rm 3D}$Barolo estimates the centre based on the total H$\alpha$ intensity map. The parameters being fitted are the position angle, the systemic velocity ($V_{\rm sys}$), the rotational velocity ($V_{{\rm rot,H}\alpha}$), and the gas velocity dispersion ($\sigma_{\mathrm{H}\alpha}$). The position angle is given an initial value based on the \textsc{galfit} measurement and allowed to vary within its uncertainties (regularised by a straight line; see \citealt{barolo}) to account for small misalignments arising from different spatial resolutions and PSFs. While we specify the redshift, we consider $V_{\rm sys}$ a free parameter to account for differences between the values reported by the KROSS and KMOS$^{\rm 3D}$ teams (based on Gaussian fits to the emission lines) and the more sophisticated kinematic $V_{\rm sys}$, which can be aggravated by the relatively low spectral resolution of the data and the noise. Finally, $\sigma_{\mathrm{H}\alpha}$ and especially $V_{{\rm rot,H}\alpha}$ are the main parameters we are interested in.

Before the fitting, we need to specify a few additional settings in the $^{\rm 3D}$Barolo configuration. First, the ring separation for our rotation curves. As a compromise between obtaining a good sampling of the rotation curve and avoiding strong correlations between pixels, we adopt a ring separation of FWHM/2, where FWHM is the reported FWHM of the PSF by the KROSS and KMOS$^{\rm 3D}$ collaborations. Second, there is the mask: we follow \cite{paper_galaxyhalo} and derive first an automatic mask using the \texttt{SEARCH} task of $^{\rm 3D}$Barolo, which we then enlarge spatially by 2 pixels; this allows us to include low S/N regions while minimising the contribution from the noisy background. With this, we proceed to obtain our kinematic models using a $\cos(\theta)$ weighting scheme and azimuthal normalisation (see \citealt{barolo} and its documentation for details). We have built mock data cubes that mimic KMOS IFU data and find that the intrinsic $V_{{\rm rot,H}\alpha}$ and $\sigma_{\mathrm{H}\alpha}$ are well recovered using this scheme.

\begin{figure}[h]
    \includegraphics[width=1\linewidth]{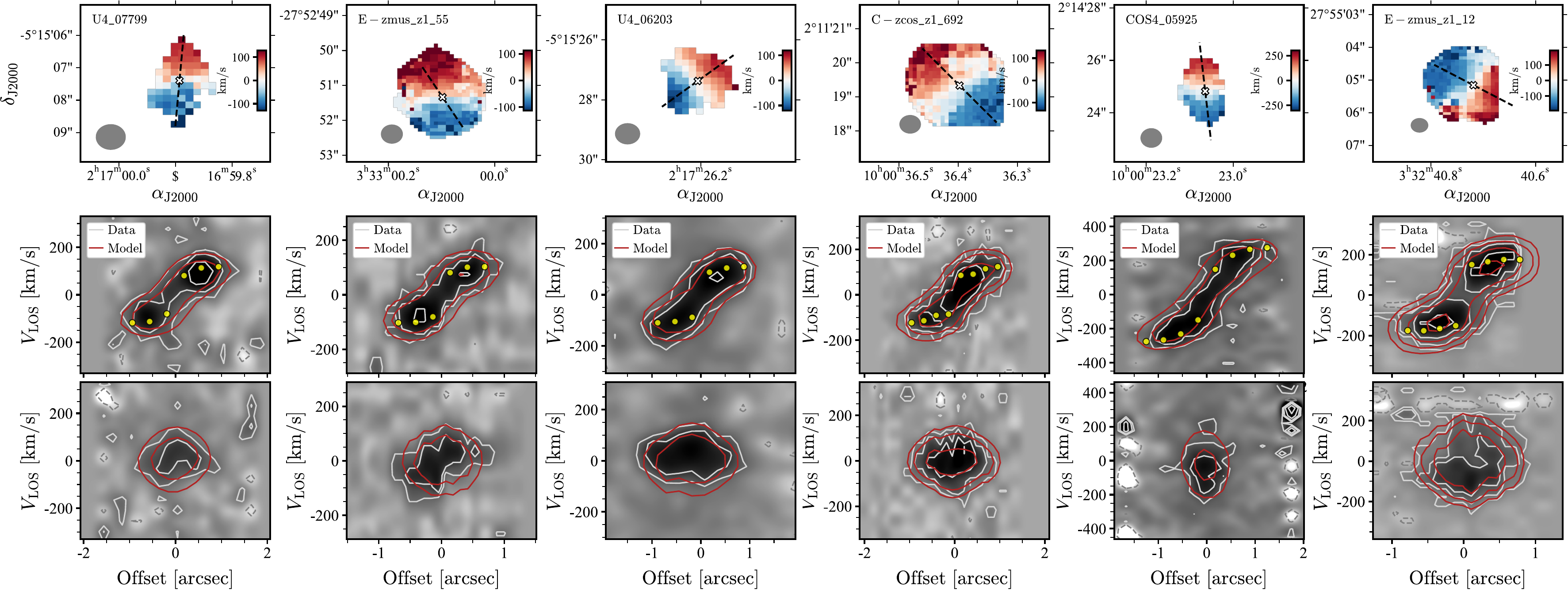}
    \caption{Examples of the kinematic data and models for our sample across our $M_\ast$ range. We show their velocity field (\emph{top}), major-axis PV (\emph{middle}), and minor-axis PV (\emph{bottom}). The top panels also show the centre (white cross), kinematic major axis (dashed black line), and PSF (grey ellipse). In the PV plots, data are shown in a grey background and black contours (grey for negative values), and the best-fitting $^{\rm 3D}$Barolo model is in red. Line-of-sight velocities are shown in yellow. Contours are at $-2,2,2^n\sigma_{\rm r.m.s.}$.}
    \label{fig:pvs}
\end{figure}

\begin{figure}
    \centering
    \includegraphics[width=0.95\linewidth]{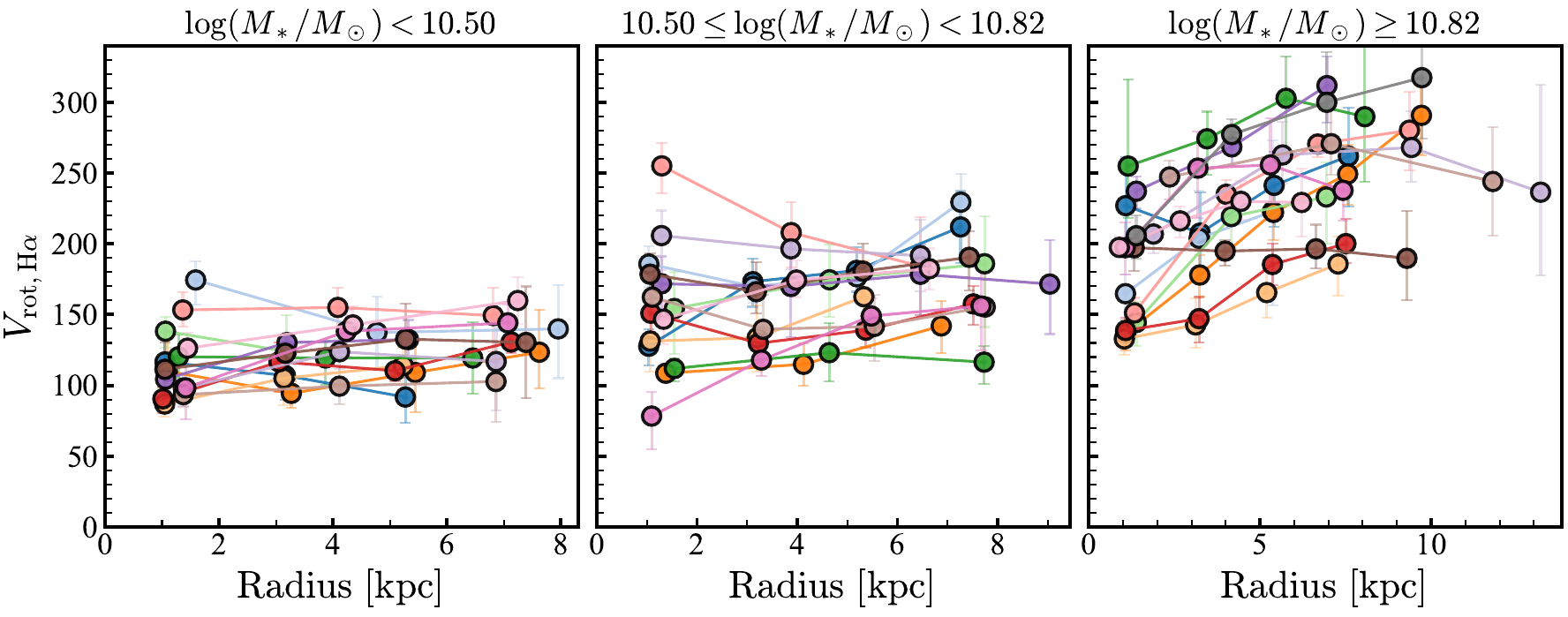}
    \caption{H$\alpha$ rotation curves of our galaxy sample, derived with $^{\rm 3D}$Barolo. Galaxies are divided into three bins according to their stellar mass. All our rotation curves and velocity dispersion measurements are available upon request.}
    \label{fig:rotcurves}
\end{figure}

After excluding galaxies for which reliable kinematic modelling could not be obtained (e.g. when emission across channel maps cannot be reproduced, PV diagrams do not indicate a rotating disc, or the data are strongly asymmetric; see \citealt{rizzo2022}), our final sample consists of 43 disc galaxies with robust kinematic models. Figure~\ref{fig:pvs} presents six representative examples of our kinematic data and models across the $M_\ast$ range, while Figure~\ref{fig:rotcurves} shows the corresponding rotation curves. The sample is composed of regularly rotating, rotation-dominated discs, with a median $\sigma_{{\rm med,H}\alpha}$ of 40 km/s and a median $V_{\rm f}/\sigma_{{\rm med,H}\alpha}$ of 5 (Fig.~\ref{fig:sample}). 

\begin{figure}[h]
    \begin{minipage}{0.73\linewidth}
        \includegraphics[width=\linewidth]{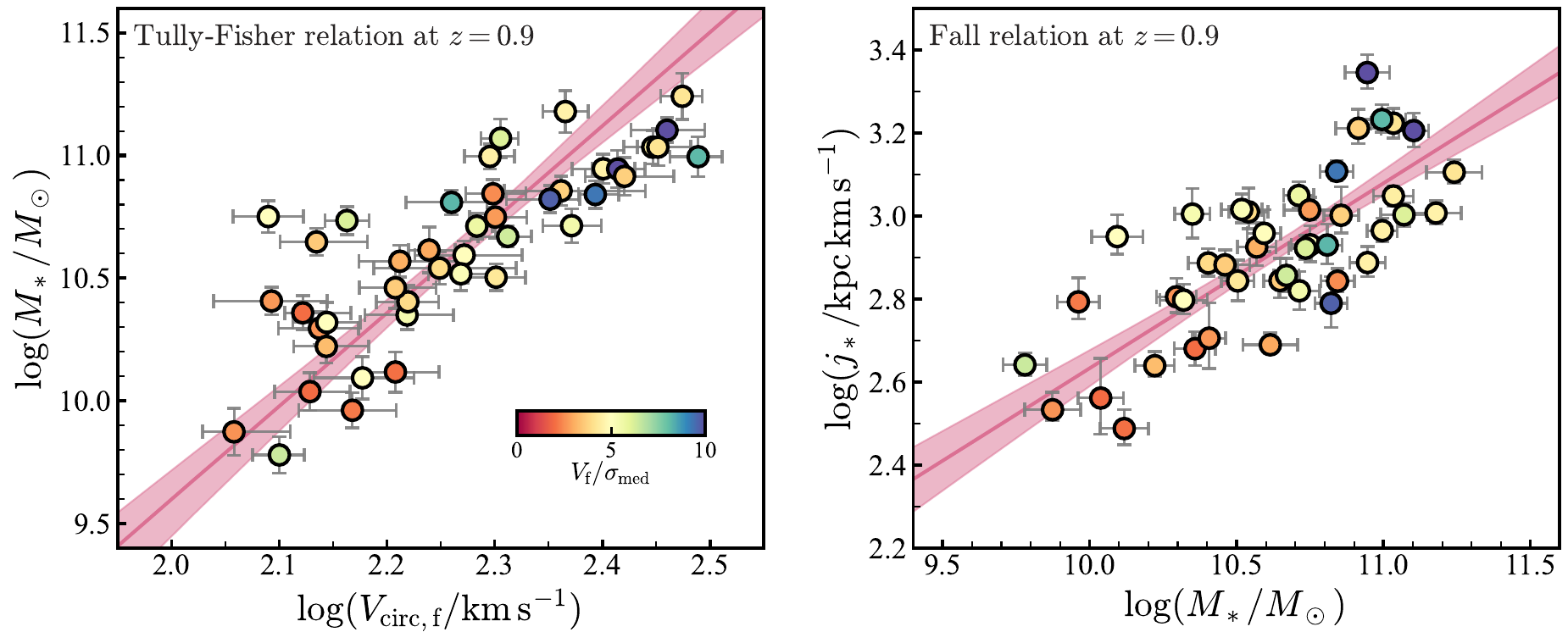}
    \end{minipage}%
    \hspace{0.03\linewidth}%
    \begin{minipage}{0.22\linewidth}
        \captionof{figure}{Dependency of the $z=0.9$ TFR and FR on $V_{\rm f}/\sigma_{\rm med}$. The solid pink curves and shaded bands represent our best-fit relations and their $1\,\sigma$ confidence bands, respectively.}
        \label{fig:j_vsigma}
    \end{minipage}
\end{figure}

\noindent
In Sect.~\ref{sec:scaling}, we present our $z=0.9$ TFR and FR. Upon close inspection of the data, we notice that there is a trend with $V_{\rm f}/\sigma_{\rm med}$, such that at fixed $M_\ast$ galaxies with higher $V_{\rm f}/\sigma_{\rm med}$ have also a higher $V_{\rm f}$ and $j_\ast$. This is shown in Fig.~\ref{fig:j_vsigma}. Similar trends for the FR at $z\sim1-3$ have been reported before (e.g. \citealt{burkert_j,kross_harrison,bouche2021}), and we have now confirmed them with accurate kinematic modelling that self-consistently derives the rotation and velocity dispersions.

\section{Calibrating our measurement technique with galaxies at $z\sim0$}
\label{app:sparc}

Given our methodology presented in Sect.~\ref{sec:method}, it is particularly crucial to ensure that we can recover the characteristic velocities at the flat part of our rotation curves (either $V_{\rm circ,f}$ or $V_{\rm f}$). To ensure that our methods are reliable, we tested and calibrated them on a sample of nearby disc galaxies (which have a similar S\'ersic distribution to our sample) using detailed, accurate measurements of $V_{\rm f}$ and $j_\ast$ based on high-resolution rotation curves and NIR photometry. Specifically, we used the SPARC compilation \citep{sparc}, which is also the sample used by \cite{marasco_mstar} to derive the $z=0$ TFR and FR we compare against in Sect.~\ref{sec:evolution}. The SPARC galaxies have $V_{\rm f}$ and $j_\ast$ measurements from \cite{sparc}, \cite{postijstar} and \cite{paperIBFR}, respectively, which were derived from the data without the use of any functional forms.

We have corroborated that our approach is robust as follows. First, we cut the SPARC rotation curves so they do not extend beyond 4 times their effective radius (derived from Sérsic fits to the SPARC surface brightness profiles). Second, we downsample the rotation curves by linear interpolation to extract rotational speeds at only four radii (the results remain unchanged when performed at three or five radii). These two steps ensure that the downsampled rotation curves resemble those from the $z=0.9$ H$\alpha$ data. Next, we fit the rotation curves with arctan profiles as in Sect.~\ref{sec:method}.

\begin{figure}[h]
    \begin{minipage}{0.7\linewidth}
        \includegraphics[width=\linewidth]{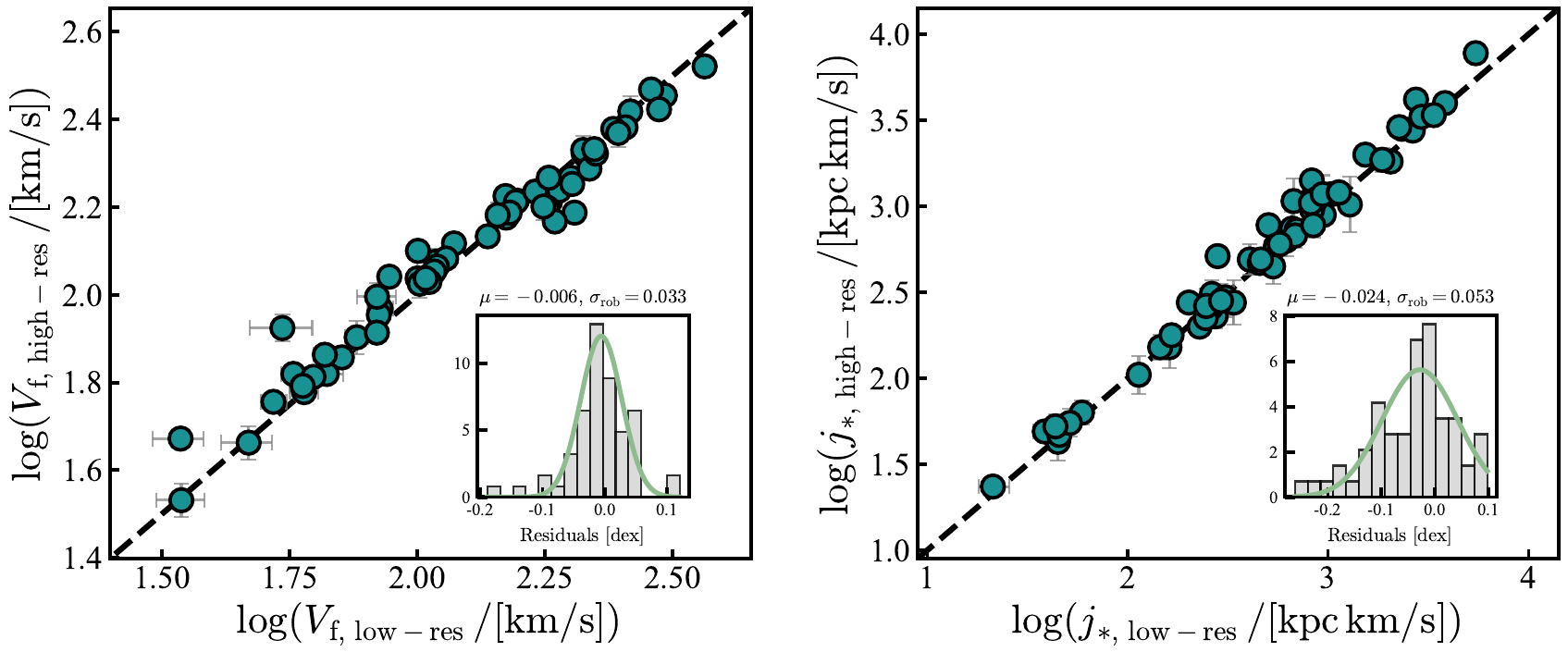}
    \end{minipage}%
    \hspace{0.02\linewidth}
    \begin{minipage}{0.26\linewidth}
        \captionof{figure}{Calibration of our $V_{\rm f}$ (\emph{left}) and $j_\ast$ (\emph{right}) estimates. The figure compares the measurements at $z=0$ obtained from high-resolution kinematics with those from downsampled rotation curves, which have a low spatial sampling similar to the $z=0.9$ H$\alpha$ data. The agreement is excellent, as highlighted by the residuals shown in the insets. }
        \label{fig:sparc}
    \end{minipage}
\end{figure}

From empirical tests, we find that the best way to retrieve the characteristic flat velocity from the low-resolution data is to evaluate the best-fitting arctan model at $R=2\,R_{\rm eff}$ for galaxies with $V_{\rm a}>100\,\rm{km/s}$ and $R=3\,R_{\rm eff}$ for galaxies with $V_{\rm a}<100\,\rm{km/s}$. As shown in the left panel of Fig.~\ref{fig:sparc}, this results in an excellent agreement with the high-resolution measurements, with only a negligible offset. Similarly, we measure $j_\ast$ sampling Eq.~\ref{eq:j} as described in the main text, and compare our values against the high-resolution estimates (right panel in Fig.~\ref{fig:sparc}). Once again, there is an excellent agreement, with low residuals and scatter. From this discussion, we conclude that our methodology for deriving the characteristic flat velocity and $j_\ast$ is robust. We have also double-checked that the $z=0$ TFR and FR re-derived with this technique have slopes and intercepts compatible with those inferred by \cite{marasco_mstar}.

\section{Abundance-matching expectations}
\label{app:AM}

\begin{figure}[h]
    \begin{minipage}{0.71\linewidth}
        \includegraphics[width=1\linewidth]{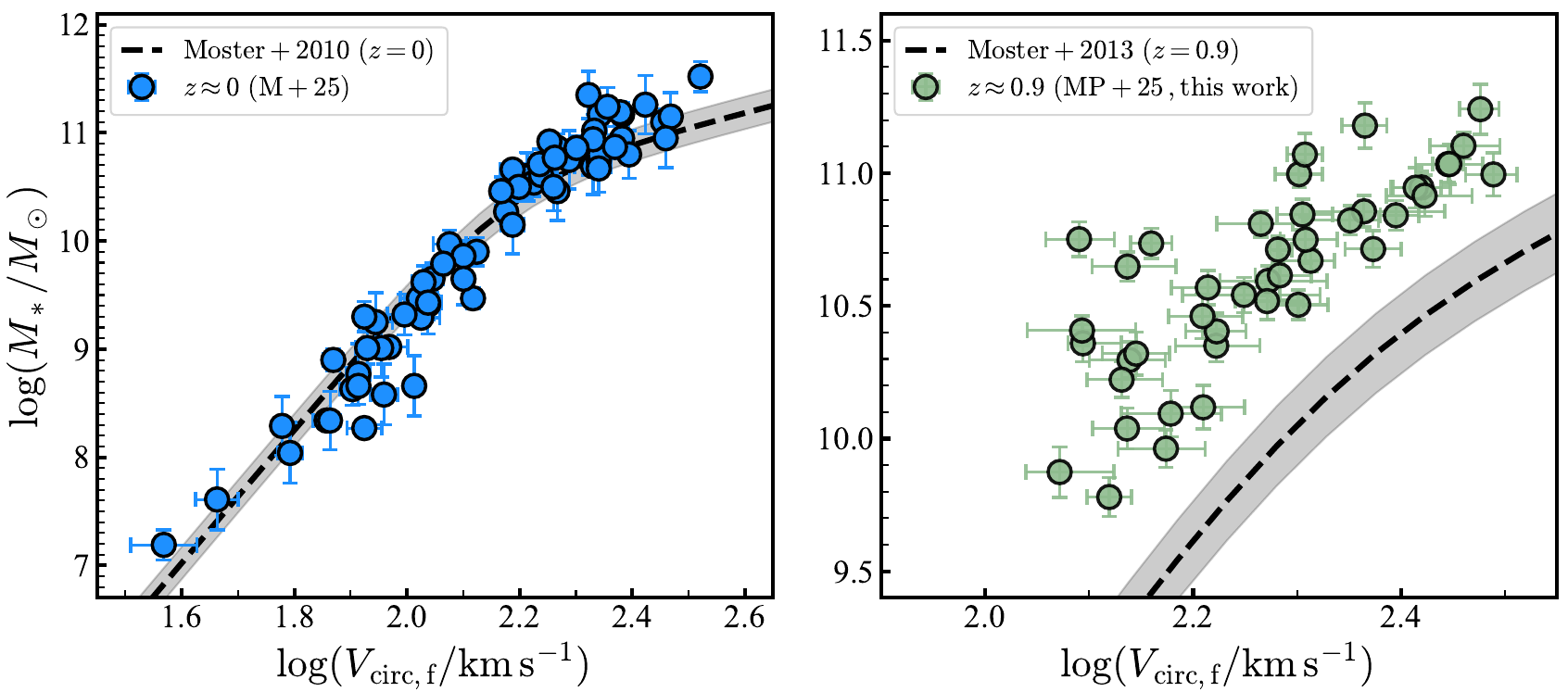}
    \end{minipage}%
    \hspace{0.02\linewidth}%
    \begin{minipage}{0.25\linewidth}
        \captionof{figure}{Observed TFRs against the predictions from abundance matching, after assuming the $f_{\rm V}$ discussed in the main text. Note the differences in the axes span.}
        \label{fig:am}
    \end{minipage}
\end{figure}

As discussed in the main text, the $f_{\rm M}$ values found at $z=0.9$ (but also at $z=0$) are different from those expected from \cite{moster2010} based on abundance-matching techniques. To visualise the level of this disagreement, in Fig.~\ref{fig:am} we compare the observed TFRs with those implied by the $f_{\rm M}-M_\ast$ relations from \cite{moster2010}, after assuming the same $f_{\rm V}$ as in Sect.~\ref{sec:origin}, i.e. solving Eq.~\ref{eq:fm} for the abundance-matching TFRs.

At low $z$ (note that here we have assumed $V_{\rm circ,f}=V_{\rm f}$) we can see discrepancies previously known in the literature, namely some dwarfs with lower $M_\ast$ than expected (the `baryon deficient dwarfs' from \citealt{paper_galaxyhalo}, see also \citealt{forbes2024}), and some massive spirals with higher $M_\ast$ than predicted by abundance matching (e.g. \citealt{postinomissing,marasco_DMinsim,enrico_massmodels_ss,paper_massmodels}). Our observations show that the discrepancies with abundance matching become much more dramatic at $z=0.9$, with the abundance-matching TFR being about 0.7 dex too low in $\log(M_\ast)$ or $0.25$ dex in $\log(V_{\rm f})$, and with a curvature not evident in the data. Alternatively, setting $f_{\rm V}=0.8$ would make the abundance-matching expectations consistent with the data, but it is unlikely that such a low value is valid for our $M_\ast$ range \citep{mcgaugh_mondBTFR}. 

\newpage
\section{Sample and main physical parameters}

\begin{table}[H]
\caption{Main parameters of our galaxy sample.}
\label{tab:cat}
\centering
\begin{tabular}{lcccccccccc}
	\hline \noalign{\vskip 1.1pt} 
 Name & $z$   & $\log(M_\ast/M_\odot)$ & $\delta_{\log(M_\ast/M_\odot)}$ & \multicolumn{3}{c}{$j_\ast$} & \multicolumn{3}{c}{$V_{\rm circ,f}$} & $V_{\rm f}/\sigma_{{\rm med,H}\alpha}$  \\ \noalign{\vskip 1.5pt}
      &  &  & &  p16 & p50 & p84 & p16 & p50 & p84 &\\ \noalign{\vskip 1.1pt}
      (1) & (2) & (3) & (4) &  (5) & (6) & (7) & (8) & (9) & (10) & (11)\\ \noalign{\vskip 1.1pt}
   \hline \noalign{\vskip 1.5pt}  
\noalign{\vskip 1.1pt}
C-HiZ_z1_245   &    0.833  & 10.50  &  0.06  &  620.17  &  697.33  &  778.75 &  188.02 &  200.13 &  212.63 &  4.3 \\ \noalign{\vskip 1.1pt}
C-HiZ_z1_258   &    0.838  & 10.71  &  0.05  & 1043.03  & 1119.82  & 1204.33 &  188.77 &  192.06 &  198.48 &  5.7 \\ \noalign{\vskip 1.1pt}
C-HiZ_z1_289   &    0.843  & 10.65  &  0.06  &  632.10  &  698.67  &  785.08 &  125.69 &  136.34 &  151.19 &  3.6 \\ \noalign{\vskip 1.1pt}
C-zcos_z1_182  &    0.970  & 10.95  &  0.06  &  711.37  &  771.99  &  841.24 &  234.93 &  251.63 &  268.57 &  4.6 \\ \noalign{\vskip 1.1pt}
C-zcos_z1_188  &    0.880  & 10.04  &  0.08  &  291.42  &  364.33  &  444.86 &  124.25 &  134.46 &  145.98 &  2.0 \\ \noalign{\vskip 1.1pt}
C-zcos_z1_189  &    0.938  & 10.86  &  0.06  &  905.49  & 1003.39  & 1162.25 &  202.03 &  229.95 &  271.30 &  3.7 \\ \noalign{\vskip 1.1pt}
C-zcos_z1_192  &    0.914  & 10.57  &  0.07  &  754.47  &  841.57  &  955.04 &  148.65 &  162.92 &  183.53 &  3.1 \\ \noalign{\vskip 1.1pt}
C-zcos_z1_198  &    0.959  & 10.84  &  0.06  & 1215.18  & 1280.64  & 1355.75 &  238.55 &  247.41 &  259.88 &  9.3 \\ \noalign{\vskip 1.1pt}
C-zcos_z1_202  &    0.841  & 10.84  &  0.06  &  657.38  &  696.55  &  745.25 &  187.24 &  198.79 &  212.85 &  2.5 \\ \noalign{\vskip 1.1pt}
C-zcos_z1_484  &    0.855  & 11.10  &  0.05  & 1460.18  & 1604.43  & 1759.81 &  266.29 &  288.37 &  311.64 & 12.4 \\ \noalign{\vskip 1.1pt}
C-zcos_z1_5    &    0.869  & 10.82  &  0.06  &  533.55  &  616.09  &  707.02 &  203.68 &  224.60 &  238.76 &  9.7 \\ \noalign{\vskip 1.1pt}
C-zcos_z1_611  &    0.835  & 10.75  &  0.07  &  770.92  &  851.95  &  942.51 &  113.74 &  122.98 &  132.21 &  5.0 \\ \noalign{\vskip 1.1pt}
C-zcos_z1_619  &    0.838  & 10.59  &  0.06  &  816.49  &  907.99  & 1026.06 &  167.18 &  186.92 &  210.09 &  5.0 \\ \noalign{\vskip 1.1pt}
C-zcos_z1_621  &    0.919  & 10.35  &  0.06  &  876.00  & 1010.13  & 1154.16 &  150.52 &  165.46 &  181.88 &  5.4 \\ \noalign{\vskip 1.1pt}
C-zcos_z1_659  &    0.982  & 10.36  &  0.07  &  434.12  &  478.96  &  523.89 &  127.30 &  132.39 &  145.95 &  2.0 \\ \noalign{\vskip 1.1pt}
C-zcos_z1_660  &    0.929  & 10.73  &  0.06  &  783.93  &  836.18  &  894.51 &  138.66 &  145.50 &  152.36 &  6.2 \\ \noalign{\vskip 1.1pt}
C-zcos_z1_690  &    0.925  & 10.71  &  0.07  &  591.24  &  660.30  &  730.56 &  220.71 &  235.15 &  250.11 &  5.5 \\ \noalign{\vskip 1.1pt}
C-zcos_z1_692  &    0.927  & 11.00  &  0.05  &  864.57  &  921.74  &  985.93 &  186.75 &  197.53 &  207.85 &  4.5 \\ \noalign{\vskip 1.1pt}
C-zcos_z1_698  &    0.927  & 10.81  &  0.05  &  761.27  &  850.90  &  949.66 &  164.26 &  181.94 &  195.74 &  8.2 \\ \noalign{\vskip 1.1pt}
C-zcos_z1_726  &    0.928  & 10.67  &  0.04  &  654.08  &  718.95  &  786.09 &  192.24 &  205.15 &  215.64 &  6.4 \\ \noalign{\vskip 1.1pt}
C-zcos_z1_793  &    0.885  & 10.41  &  0.06  &  421.85  &  507.93  &  606.71 &  108.54 &  123.83 &  138.58 &  2.7 \\ \noalign{\vskip 1.1pt}
COS4_05925     &    0.799  & 11.03  &  0.07  & 1056.84  & 1117.08  & 1178.54 &  270.52 &  279.69 &  289.03 &  4.5 \\ \noalign{\vskip 1.1pt}
COS4_06487     &    0.907  & 11.24  &  0.09  & 1194.21  & 1274.12  & 1362.13 &  284.26 &  297.95 &  310.59 &  4.2 \\ \noalign{\vskip 1.1pt}
COS4_16172     &    1.030  & 11.03  &  0.08  & 1532.45  & 1671.95  & 1807.81 &  262.11 &  282.77 &  302.68 &  4.2 \\ \noalign{\vskip 1.1pt}
COS4_17628     &    0.907  & 11.07  &  0.08  &  942.80  & 1005.94  & 1073.15 &  193.59 &  202.04 &  209.76 &  6.2 \\ \noalign{\vskip 1.1pt}
COS4_23890     &    0.852  & 10.95  &  0.08  & 2016.41  & 2213.38  & 2431.61 &  245.49 &  259.50 &  274.59 & 12.7 \\ \noalign{\vskip 1.1pt}
C-zcos_z1_675  &    0.891  & 10.91  &  0.08  & 1488.39  & 1625.61  & 1798.41 &  241.26 &  263.33 &  291.14 &  3.7 \\ \noalign{\vskip 1.1pt}
C-zcos_z1_693  &    0.925  & 10.30  &  0.07  &  583.29  &  638.75  &  703.67 &  125.15 &  137.19 &  148.70 &  2.6 \\ \noalign{\vskip 1.1pt}
E-zmus_z1_119  &    0.841  & 10.75  &  0.09  &  978.12  & 1032.94  & 1104.22 &  188.77 &  199.82 &  213.29 &  2.7 \\ \noalign{\vskip 1.1pt}
E-zmus_z1_12   &    0.976  & 11.18  &  0.09  &  956.32  & 1015.91  & 1085.51 &  220.99 &  232.06 &  243.44 &  4.6 \\ \noalign{\vskip 1.1pt}
E-zmus_z1_148  &    0.840  & 10.12  &  0.08  &  280.25  &  307.62  &  340.04 &  150.02 &  161.39 &  176.51 &  2.1 \\ \noalign{\vskip 1.1pt}
E-zmus_z1_166  &    0.975  & 10.62  &  0.09  &  465.55  &  489.33  &  521.80 &  165.74 &  173.41 &  183.11 &  3.0 \\ \noalign{\vskip 1.1pt}
E-zmus_z1_202  &    0.861  &  9.78  &  0.07  &  412.20  &  437.90  &  466.65 &  118.77 &  125.95 &  132.66 &  6.5 \\ \noalign{\vskip 1.1pt}
E-zmus_z1_24   &    0.823  & 10.40  &  0.07  &  713.44  &  770.29  &  831.07 &  154.63 &  165.75 &  176.59 &  3.9 \\ \noalign{\vskip 1.1pt}
E-zmus_z1_49   &    0.982  & 10.54  &  0.07  &  927.67  & 1022.75  & 1156.29 &  152.38 &  177.39 &  206.40 &  3.9 \\ \noalign{\vskip 1.1pt}
E-zmus_z1_55   &    0.885  & 10.32  &  0.08  &  578.56  &  627.25  &  682.82 &  129.22 &  139.36 &  149.34 &  5.0 \\ \noalign{\vskip 1.1pt}
E-zmus_z1_65   &    0.842  &  9.96  &  0.07  &  561.49  &  621.02  &  704.09 &  130.39 &  147.22 &  161.00 &  2.2 \\ \noalign{\vskip 1.1pt}
E-zmus_z1_70   &    0.858  & 10.22  &  0.07  &  410.94  &  435.92  &  470.45 &  129.53 &  139.24 &  151.83 &  3.3 \\ \noalign{\vskip 1.1pt}
E-zmus_z1_86   &    0.841  &  9.87  &  0.10  &  320.72  &  341.47  &  374.75 &  106.57 &  114.30 &  128.05 &  2.6 \\ \noalign{\vskip 1.1pt}
U4_06203       &    0.803  & 10.52  &  0.07  &  960.35  & 1036.43  & 1125.88 &  172.68 &  185.59 &  201.76 &  5.2 \\ \noalign{\vskip 1.1pt}
U4_07799       &    0.809  & 10.09  &  0.09  &  804.60  &  891.60  & 1000.24 &  134.90 &  150.42 &  166.94 &  5.0 \\ \noalign{\vskip 1.1pt}
U4_22990       &    0.787  & 10.99  &  0.08  & 1585.61  & 1706.51  & 1847.94 &  289.62 &  308.19 &  323.95 &  8.1 \\ \noalign{\vskip 1.1pt}
U4_34173       &    0.896  & 10.46  &  0.08  &  709.28  &  762.21  &  825.37 &  148.93 &  161.36 &  175.47 &  3.6 \\ \noalign{\vskip 1.1pt}
   \hline
    \end{tabular} 
    \tablefoot{Column (1) gives the galaxy ID as provided by the KROSS and KMOS$^{\rm 3D}$ teams. Column (2) is the redshift from the KROSS and KMOS$^{\rm 3D}$ catalogues. Columns (3) and (4) quote the stellar masses and their uncertainties. Columns (5) to (7) give the 16th, 50th, and 84th percentiles of our $j_\ast$ estimates. Columns (8) to (10) give the 16th, 50th, and 84th percentiles of $V_{\rm circ,f}$. Column (11) lists the rotation to dispersion ratio $V_{\rm f}/\sigma_{{\rm med,H}\alpha}$.}
\end{table}

\end{appendix}

\end{document}